\documentclass[twocolumn,nolinenumbers,trackchanges]{aastex701}
\usepackage{amsmath}
\usepackage{amssymb}
\usepackage{amsthm}
\usepackage{natbib,aasdefs,url,bm}
\usepackage{array}
\usepackage{float}
\usepackage{graphicx}
\usepackage{subfigure}
\usepackage[dvipsnames]{xcolor}
\hypersetup{linkcolor=red,citecolor=NavyBlue,filecolor=cyan,urlcolor=purple}

\def\be{\begin{equation}}
\def\ee{\end{equation}}
\def\ba{\begin{eqnarray}}
\def\ea{\end{eqnarray}}

\def\aap{\ {A\&A}\ }

\def\apjl{\ {ApJL}\ }
\def\apjs{\ {ApJS}\ }

\shorttitle{High energy neutrinos from pulsar-powered optical transients}
\shortauthors{Mukhopadhyay and Kimura}
\begin{document}
\title{High energy neutrinos from pulsar-powered optical transients: LFBOTs as potential origin of the KM3NeT event KM3-230213A}
\correspondingauthor{Mainak Mukhopadhyay}
\email{mainak@fnal.gov}
\author[0000-0002-2109-5315]{Mainak Mukhopadhyay}
\affiliation{Astrophysics Theory Department, Theory Division, Fermi National Accelerator Laboratory, Batavia, Illinois 60510, USA}
\altaffiliation{FERMILAB-PUB-26-0014-T}
\affiliation{Kavli Institute for Cosmological Physics, University of Chicago, Chicago, Illinois 60637, USA}
\email{mainak@fnal.gov}
\author[0000-0003-2579-7266]{Shigeo S. Kimura}
\affiliation{Frontier Research Institute for Interdisciplinary Sciences; Astronomical Institute, Graduate School of Science, Tohoku University, Sendai 980-8578, Japan}
\email{shigeo@astr.tohoku.ac.jp}
\begin{abstract}
Recently, the KM3NeT Collaboration reported the detection of an ultra-high energy ($\sim 220$ PeV) neutrino event, KM3-230213A. In this work, we perform a detailed investigation into whether this event could originate from the diffuse neutrino flux produced by a class of pulsar-powered optical transients. In particular, we consider populations of ordinary supernovae (SNe), super-luminous supernovae (SLSNe), and luminous fast blue optical transients (LFBOTs) with a newly formed magnetar as the central engine. We discuss both the thermal electromagnetic and non-thermal neutrino emission from such sources. We scan the parameter space of the dipolar magnetic field strength and the initial spin period to determine characteristic optical emission properties and lightcurve timescales of these transients. Additionally, our scan identifies which classes of these transients can reproduce the required diffuse flux level and neutrino energies. Combining our results, we conclude that a diffuse neutrino flux from a population of LFBOTs can explain the KM3NeT event. Therefore, pulsar-powered optical transients may serve as promising sources for the current and upcoming high-energy and ultra-high energy neutrino telescopes.
\end{abstract}
\section{Introduction}
\label{sec:intro}
The detection of ultrahigh energy (UHE) neutrinos (with $E_\nu>10^{17}$ eV) has been an ongoing pursuit in the planning and building of neutrino telescopes. These neutrinos at the highest energies can have an astrophysical or a cosmogenic origin. The former is sourced by UHE cosmic ray (UHECR) accelerators following which they interact with photons or baryons. The latter is sourced when these UHECRs escape the source and interact with the intergalactic medium during their flight from the astrophysical source to the Earth. Until very recently, there were limits on the plausible flux of such neutrinos from various ongoing experiments like IceCube~\citep{IceCubeCollaborationSS:2025jbi} and Auger~\citep{PierreAuger:2023pjg}. However, the recent discovery of an UHE neutrino event at KM3NeT~\citep{KM3NeT:2025npi} has bolstered both the search and modeling of UHE neutrino sources (or flux) across all frontiers.

The KM3NeT event, KM3-230213A, was detected by the Astroparticle Research with Cosmics in the Abyss (ARCA) detector (with 21 operational strings and 287.4 days of operation) of KM3NeT in the Mediterranean Sea on February 13, 2023. The reconstructed energy of the muon corresponding to the track topology was estimated to be $120^{+110}_{-60}$ PeV at 68\% C.L. This implies that the parent neutrino energy assuming $dN/dE_\nu \propto E_\nu^{-2}$ is roughly $220$ PeV with error bands being at $110 - 790$ PeV at 68\% and $72\ {\rm PeV} - 2.6\ {\rm EeV}$ at 90\% C.L. Assuming an isotropic $E_\nu^{-2}$ UHE neutrino flux, a joint fit to the KM3NeT event and the non-observations from IceCube and Pierre Auger observatories provides a diffuse flux normalization that reconciles the KM3NeT event with data from IceCube and Auger~\citep{KM3NeT:2025ccp}. We refer to this as the \emph{joint flux} (of diffuse neutrinos at UHE) from here on.

Several speculations regarding the origin of this event have been made invoking sources both within the standard model and beyond. The Standard model explanations include astrophysical origins from sources like blazars or active galactic nuclei~\citep{Dzhatdoev:2025sdi,KM3NeT:2025bxl,KM3NeT:2025lly,Yuan:2025zwe,deClairfontaine:2025gei} and blazar flares~\citep{Neronov:2025jfj} and cosmogenic origins~\citep{Kuznetsov:2025ehl,Boxi:2025ony}. Due to the $\mathcal{O}(\rm EeV)$ parent CR energy, the event is believed to have an extra-galactic origin~\citep{KM3NeT:2025aps}. The beyond Standard model scenarios include decay of super-heavy dark matter~\citep{Aloisio:2025nts,Borah:2025igh,Kohri:2025bsn,Khan:2025gxs,Murase:2025uwv,Narita:2025udw}, boosted dark matter~\citep{Dev:2025czz}, and evaporation of primordial black holes~\citep{Boccia:2025hpm,Airoldi:2025opo}. 

In this work, we aim to investigate the possibility of an astrophysical origin of KM3-230213A, in particular, as a result of a diffuse neutrino flux from a class of optical transients. Core-collapse events can produce highly-magnetized neutron stars (dipolar field $B_d \sim 10^{13} - 10^{15}$ G) with millisecond spin periods ($P_i \sim 1 - 10$ ms), whose rotational energy $E_{\rm rot} \sim 2 \times 10^{52}\ {\rm erg}\ (P_i/1\ {\rm ms})^{-2}$ serves as a powerful energy reservoir. These pulsar-powered transients span a range of observed populations, from ordinary SNe to super-luminous supernovae (SLSNe), and luminous fast blue optical transients (LFBOTs), with their optical lightcurve evolution primarily governed by the spindown timescale $t_{\rm sd}\sim 5.6 \times 10^{5}\ {\rm s}\ (B_d/10^{14}G)^{-2} (P_i/3\ \rm ms)^{2}$ relative to photon diffusion through the ejecta. 

Pulsars embedded in supernova ejecta can source UHECRs that subsequently produce high energy neutrinos through hadronuclear ($pp$) and photohadronic ($p\gamma$) interactions~\citep{Horiuchi:2008zc, Murase:2009pg}, with emission timescales ranging from days to months post-collapse depending on ejecta properties~\citep{Fang:2013vla,Fang:2014qva,Kashiyama:2015eua}. The large spindown energies in SLSNe can further enhance neutrino production~\citep{Fang:2015xhg,Prepration:2026slsne}, while LFBOTs like AT2018cow~\citep{Prentice:2018qxn}, with their powerful central engines and low-mass ejecta, may produce intense but short-lived neutrino signals~\citep{Fang:2018hjp,Guarini:2022uyp}.

The central question we want to address in this \emph{letter} is whether a diffuse high energy neutrino flux from a class of optical transients with a new born pulsar as the central engine can be a plausible explanation for the KM3-230213A. Although we deal with this central question, the current work is broad in scope, where we survey the parameter space associated with two important physical properties of the pulsar- the initial spin period $P_i$ and dipolar magnetic field strength $B_d$. We identify three distinct classes of optical transients powered by pulsar remnants -- ordinary SNe, SLSNe, and LFBOTs. We find that LFBOTs are ideally suited to produce the neutrino energy and fluence needed for the joint flux. This not only suitably explains the properties of the KM3NeT event, but is also timely given the interest in UHE neutrino searches and the promising progress of optical transient astronomy.

The \emph{Letter} is organized as follows. The model we use for computing the electromagnetic and neutrino emission is briefly summarized in Section.\ref{sec:model}. In Section~\ref{sec:lfbot} we use our model and present a chosen scenario (the one that contributes the maximum to the diffuse flux) of neutrino and EM emissions from a LFBOT. The survey of the parameter space for the possible pulsar powered optical transients is outlined in Section~\ref{sec:survey}. The diffuse contribution from LFBOTs and the results successfully explaining a plausible origin of the KM3NeT event is discussed in Section~\ref{sec:diffuse}. We summarize and discuss the implications of our work in Section~\ref{sec:disc}.
\section{Model overview}
\label{sec:model}
In this section, we briefly discuss the model of the central engine that powers the optical transients and their corresponding surrounding environment. As mentioned above, the central engine of all the transients discussed in this work are newly-born millisecond pulsars (or magnetars). The model we use for this work is discussed in detail in~\cite{Mukhopadhyay:2024ehs} and~\cite{Mukhopadhyay:2025tvz}. In essence, we have the pulsar which deposits its spindown energy in the surrounding nebula. The nebula is surrounded by an ejecta whose properties depend on the class of transients. 

The spindown luminosity of a millisecond pulsar ($L_{\rm sd}$) is $\propto B_d^2 P_i^{-4} \left( 1+t/t_{\rm sd} \right)^{-2}$, where $B_d$ is the dipolar magnetic field strength, $P_i$ is the initial spin period, and $t_{\rm sd}$ is the spindown timescale which is $\propto B_d^{-2} P_i^{2}$. For this work, we assume rapidly spinning pulsars with $P_i$ ranging from $1 - 10$ milliseconds and $B_d$ varying between $10^{12} - 10^{15}$ G. As the pulsar spins down, the spindown energy of the pulsar is converted into non-thermal ($E_{\rm nth}$), thermal ($E_{\rm th}$), magnetic ($E_B$), and kinetic energy doing $PdV$ work on the system (see Appendix~\ref{appsec:energy} for details). The pulsar loses its spin energy due to pulsar-like or magnetically-driven winds. The spindown energy deposited through shocks or magnetic reconnection aids in the formation of a hot nebula behind an expanding ejecta. Thus, physically the system consists of the pulsar, the wind region, the nebula and the ejecta.

The pulsar wind comprises of $e^+ - e^-$ pairs which are extracted from the magnetosphere through $B-\gamma$ and $\gamma-\gamma$ processes. The ultra-relativistic wind encounters the expanding ejecta, decelerates, and thermalizes to form a termination shock (TS).
The shocked region downstream of the TS, leads to the formation of a hot, magnetized nebula which is confined by the supernova ejecta. The nebula is filled with $e^+ - e^-$ pairs and non-thermal photons (radiation dominated). While the former is accounted for by the pulsar wind and pair production by the photons, the latter is sourced through non-thermal emission from the $e^+ - e^-$ pairs. The injected $e^+ - e^-$ pairs follow a broken power law similar to~\cite{Mukhopadhyay:2025tvz}, $dN/d\gamma_e \sim \gamma_e^{-1.5}$ for $\gamma_m \leq \gamma_e \leq \gamma_{e,
\rm br}$ and $dN/d\gamma_e \sim \gamma_e^{-2.5}$ for $\gamma_{e,
\rm br} < \gamma_e \leq \gamma_M$, where the electron Lorentz factor $\gamma_e = \varepsilon_e/(m_e c^2)$, $\varepsilon_e$ is the comoving energy of the electron, $\gamma_{e,\rm br} = 10^3$ is the break Lorentz factor, $\gamma_M$ and $\gamma_m$ are the maximum and minimum electron Lorentz factors. 

The non-thermal photons are reprocessed into thermal photons by the ejecta. The fraction of non-thermal radiation that is reprocessed is governed by the quantity $(1-\mathcal{A})$ (where $\mathcal{A}$ is the albedo of the ejecta). We consistently compute this albedo depending on the composition of the ejecta, its opacity, and ionization state. The thermal photon spectra is given by a blackbody spectra with temperature of the thermal photons estimated using $T_{\rm th} \sim \left( E_{\rm th}/( a\ 4 \pi R^2 \Gamma_{\rm ej}^2 \beta_{\rm ej} c t ) \right)^{1/4}$ where $a$ is the radiation constant, $R$ is the effective nebula-ejecta radius, $\beta_{\rm ej}$ and $\Gamma_{\rm ej}$ are the velocity of the ejecta normalized to speed of light $c$ and the corresponding Lorentz factor respectively. 

For our models, the thermal lightcurve resulting from the thermal photons dominates over the non-thermal lightcurve in the optical band. This can be understood by the fact that the mass and composition of supernova ejecta is such that the ejecta is opaque to optical photons at times when spindown energy is deposited in the nebula. This results in efficient thermalization, where the energy density in non-thermal photons is converted to thermal photons through reprocessing in the ejecta. Furthermore, the photons suffer Synchrotron Self-Absorption (SSA) and $\gamma \gamma$ attenuation in the nebula. The ejecta comprising of various elements also contribute to attenuation in the observed EM spectrum. The details about the ejecta mass and composition for LFBOTs are discussed in Section~\ref{sec:lfbot} while that for ordinary SNe and SLSNe are detailed in Section~\ref{sec:survey}.

The co-rotation of the plasma around the pulsar leads to the creation of an electric field which can extract charges from the surface of the pulsar. The maximum charge density that can be extracted known as the Goldreich Julian (GJ) charge density is given by $n_{\rm GJ} = - \bm \Omega\cdot\bm B/\big(2 \pi Z e c\big)$~\citep{Goldreich:1969sb}, where $Z$ is the atomic number, $e$ is the charge of an electron. For simplicity, we focus on protons and have $Z = 1$. We normalize the injection spectrum for the CR protons with $n_{\rm GJ}$ which are then accelerated. We consider two primary sources of particle acceleration in the system - the polar cap region of the pulsar~\citep{Arons:2002yj,Blasi:2000xm,Fang:2013vla} and the termination shock region~\citep{Sironi:2011zf,Sironi:2014jfa}. The potential gap created in the polar cap region offers a suitable site for acceleration of the protons to high energies of $\mathcal{O}(10^9)$ GeV and above, while the TS region can further reaccelerate some of the protons through shocks. The accelerated protons interact with the non-thermal and thermal photons in the nebula and a part of them undergo photomeson cooling to produce high energy neutrinos.
\section{LFBOTs as sources of high-energy neutrinos}
\label{sec:lfbot}
\begin{figure}
\centering
\includegraphics[width=0.48\textwidth]{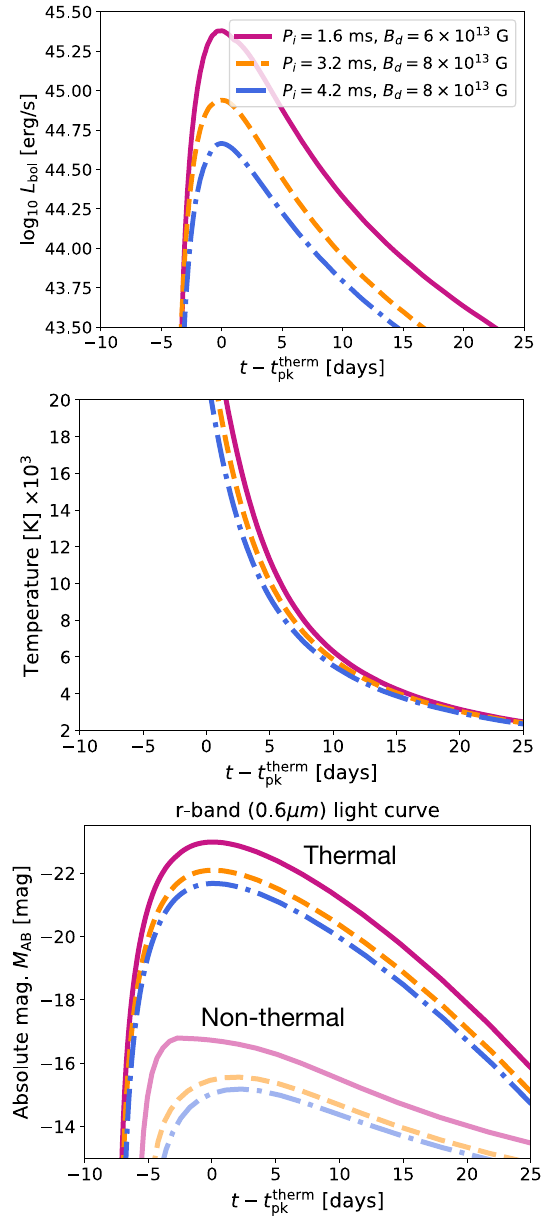}
\caption{\label{fig:lbol_temp_lc} Time evolution of the bolometric luminosity ($L_{\rm bol}$) \emph{(top)}, thermal temperature ($T_{\rm th}$) \emph{(middle)}, and thermal (darker) and non-thermal (lighter) optical lightcurves in the $r$-band ($0.6\ \mu m$) \emph{(bottom)} for three different parameter sets for LFBOTs. The time at which the bolometric luminosity peaks is defined as $t_{\rm pk}^{\rm therm}$ and is given by $t_{\rm pk}^{\rm therm} = 3.64, 3.9, 3.76$ days for $P_i = 1.6$ ms, $B_d = 6 \times 10^{13}$ G; $P_i = 3.2$ ms, $B_d = 8 \times 10^{13}$ G; and $P_i = 4.2$ ms, $B_d = 8 \times 10^{13}$ G respectively.
}
\end{figure}
In this section, we use our model discussed in Section~\ref{sec:model} to compute the non-thermal and thermal electromagnetic signatures and high energy neutrino emission from a single LFBOT. LFBOTs are rare optical transients with high peak optical luminosities ($\sim 10^{44}\ {\rm erg\ s}^{-1}$), a hot and blue spectra in the initial phase, and rapidly evolving lightcurves ($\sim$ a few days). The discovery of AT2018cow and several other transients with similar properties, have established LFBOTs as a distinct population where a plausible explanation could be a powerful newly-born pulsar embedded in a low mass ejecta~\citep{Prentice:2018qxn,Perley:2018oky,Perley:2021ell}.

For the remnant pulsar we choose $M_* = 1.42 M_\odot$ and $R_* = 12$ km as the mass and radius respectively. Since the main results of this work aim to explain the diffuse contribution of LFBOTs to explain the KM3NeT event, we will focus on the neutrino emission from the parameter set that contributes the most towards the diffuse neutrino flux, $P_i = 1.6$ ms and $B_d = 6 \times 10^{13}$ G\footnote{How we obtain this parameter set will be discussed in Sections~\ref{sec:survey} and~\ref{sec:diffuse}.}. Besides this, we also show some example parameter sets that contribute significantly to the diffuse neutrino flux when discussing the optical signatures, namely, $P_i = 3.2$ ms and $4.2$ ms for $B_d = 8 \times 10^{13}$ G for each scenario.

Since LFBOTs are associated with fairly low mass ejecta we consider 
$M_{\rm ej}^{\rm LFBOT} = 0.1 M_\odot$. The initial kinetic energy of the ejecta is fixed at $\sim 5.1 \times 10^{51}$ erg which fixes the initial ejecta velocity to be $v_{\rm ej}^{\rm LFBOT} \approx 0.24$ c. The composition of ejecta is motivated from spectroscopic observations~\citep{Perley:2018oky,Perley:2026clz} such that we have an ejecta mostly dominated by hydrogen ($50\%$) and helium ($40\%$). The remaining $10\%$ is distributed between carbon ($3\%$), oxygen ($6\%$), and silicon ($1\%$). The ejecta opacity is chosen to be $\kappa_{\rm ej}^{\rm LFBOT} = 0.2\ {\rm cm}^2{\rm g}^{-1}$.
% such that the optical depth in the ejecta is given as $\tau_\gamma = 3 M_{\rm ej} \kappa_\gamma/\left( 4 \pi R^2 \Gamma_{\rm ej}^2\right)$. 
We refer the reader to Appendix~\ref{appsec:energy} for results on the distribution of total spindown energy and the relevant timescales associated with the system.
\subsection{Optical features}
In this subsection, we focus on the main properties associated with LFBOTs as bright optical transients with comparatively short evolution timescales. The bolometric luminosity $L_{\rm bol}$ for LFBOTs where photon diffusion is dominant can be estimated using $L_{\rm bol} \sim E_{\rm th}/t_{\rm diff}^{\rm ej}$, where $E_{\rm th}$ is the energy in thermal radiation and $t_{\rm diff}^{\rm ej}$ is the diffusion timescale of photons through the ejecta. The photon diffusion timescale through the ejecta is defined as $t_{\rm diff}^{\rm ej} \approx R/c \left( 1+\tau_{\rm diff}^{\rm ej}\right)$, where the optical depth for the ejecta is given by
\begin{equation}
\label{eq:photon_diff}
\tau_{\rm diff}^{\rm ej} = \frac{\rho_{\rm ej}^\prime R \kappa_{\rm ej}}{\Gamma_{\rm ej}} = \frac{3 M_{\rm ej} \kappa_{\rm ej}}{4 \pi R^2 \Gamma_{\rm ej}^2}\,,
\end{equation}
where $\rho_{\rm ej}^\prime = M_{\rm ej}/\left( (4/3) \pi R^3 \Gamma_{\rm ej} \right)$ is the number density of the ejecta in the fluid-rest frame. In Figure~\ref{fig:lbol_temp_lc} we show the time evolution of $L_{\rm bol}$ \emph{(top)}, $T_{\rm th}$ \emph{(middle)}, and the optical $r$-band lightcurve for both thermal and non-thermal emissions \emph{(bottom)} for our model of LFBOTs. We define the time at which the bolometric luminosity peaks as $t_{\rm pk}^{\rm therm}$, and denote the time evolution in terms of $t - t_{\rm pk}^{\rm therm}$. As is typical of LFBOTs, the time needed for the peak of $L_{\rm bol}$ to decline to $L_{\rm bol}/2$ (defined as the fall time, $t_-$) is roughly between $5 - 7$ days with a peak luminosity of $\sim 2 \times 10^{45}$ erg/s. The corresponding thermal temperature drops to a $\sim {\rm a\ few\ }1000$ K over a timescale of $\mathcal{O}(10)$ days. It is worth noting here that observations of LFBOTs reveal that even though $L_{\rm bol}$ drops rapidly, they remain blue, that is, maintain a nearly constant temperature of $\sim {\rm a\ few} \times 10^{4}$ K~\citep{Omand:2026vyl,Perley:2026clz}. This is clearly not captured in our model with an expanding ejecta where the temperature also drops rapidly on timescales of $\mathcal{O}(10)$ days.

With a low ejecta mass of $M_{\rm ej}^{\rm LFBOT} = 0.1 M_\odot$, and radius $R \sim 10^{16}$ cm at $10$ days after the supernovae, the optical depth $\tau_{\rm diff}^{\rm ej} \sim 0.1$, which allows the photons to diffuse out efficiently. Moreover, the higher initial spin period $P_i$ leads to a higher peak in $L_{\rm bol}$ since shorter $P_i$ serves as stronger central engines, with larger spindown energy deposited in the nebula.

For the lightcurves, we note that the non-thermal emission is much weaker than the thermal emission, since the energy density of thermal photons is higher than the non-thermal ones (discussed in Section~\ref{sec:model}) in the optical band. The peak $M_{AB}$ brightness in this band is $\sim -22$ to $-21$ for thermal emission while it is limited to $M_{AB} \sim -16$ to $-15$ for non-thermal emissions. Similar to the bolometric luminosity, the optical lightcurve also shows $t_{-} \sim 5 - 7$ days (in this case it is defined as where the $M_{AB}$ brightness reduces by $\sim 1$). Similar to the trend in $L_{\rm bol}$ (and as expected), the LFBOTs with a smaller $P_i$ are brighter.

\begin{figure}
\centering
\includegraphics[width=0.48\textwidth]{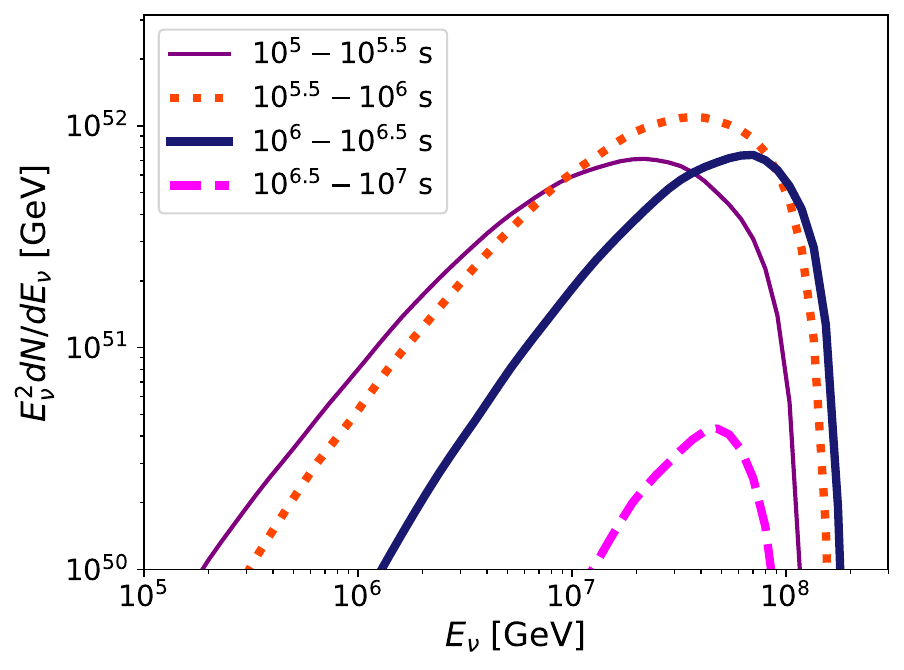}
\caption{\label{fig:nu}Total energy fluence emitted in high energy neutrinos (all flavors) from a chosen parameter set for LFBOTs ($P_i = 1.6$ ms and $B_d = 6 \times 10^{13}$ G), plotted for various time intervals ranging from a few hours to few days post the collapse.
}
\end{figure}
\subsection{Neutrino properties}
In this subsection, we discuss the high energy neutrino emission from a single LFBOT as a source. We focus on the parameter set $P_i = 1.6$ ms and $B_d = 6 \times 10^{13}$ G. The details regarding the proton injection, acceleration, maximum energy, and timescales are provided in Appendix.~\ref{appsec:proton_details}.

In Figure~\ref{fig:nu} we show the neutrino spectrum integrated over half-decades in time ranging from $10^5 - 10^7$ s. The energy in neutrinos peak between $10^{5.5} - 10^{6}$ s ($\mathcal{O}(5\ \rm days)$) post collapse. This can be explained by a combination of factors like the peak in total energy of CR protons (see Figure~\ref{fig:energy}) and their efficient cooling through the photomeson channel (see Figure~\ref{fig:nu_timescales}). Furthermore, the peak of the total energy emitted in neutrinos is roughly between $10^7 - 10^8$ GeV, an order of magnitude less than the maximum energy to which the protons are accelerated. The protons with energy $10^8$ GeV and above are efficiently cooled using the photomeson channel to produce the highest energy neutrinos ($\sim 10^7 - 10^8$ GeV). For energy $10^9$ GeV and above, although the neutrinos are efficiently produced, the acceleration of the protons is limited by polar cap acceleration, which consequently limits the maximum energy of the resulting neutrinos. For $\varepsilon_p^\prime \sim 3 \times 10^8$ GeV at $t\sim 2 \times 10^6$ s, the thermal photons ($\sim 1$ eV) serve as primary targets for the production of high energy neutrinos.

Another striking feature that can be observed is that the emission peaks between $10^{5.5} - 10^{6}$ s (dotted orange-red) but following that, it rapidly declines between $10^6 - 10^{6.5}$ s (dashed pink). This rapid decline can be explained by Figure~\ref{fig:nu_timescales} where we see that the neutrino production efficiency decreases for $t \gtrsim 10^{6.5}$ s. Furthermore, we note that $\dot{N}_p t$ is constant (see Equation~\ref{eq:prot_inj_spec}) and hence the decreases in $\varepsilon_{p}^{\prime \rm max}$ roughly dictates the trend for the energy emitted in neutrinos. From Figure~\ref{fig:epcut}, we see that between $10^{6.5} - 10^{7}$ s, $\varepsilon_{p}^{\prime \rm max}$ is dominated by the maximum energy available across the polar cap which falls off as $1/t$, explaining the rapid decline.
\section{Survey of optical transients with respect to physical parameters: $B_d - P_\MakeLowercase{i}$}
\label{sec:survey}
\begin{figure*}
\centering
\includegraphics[width=\textwidth]{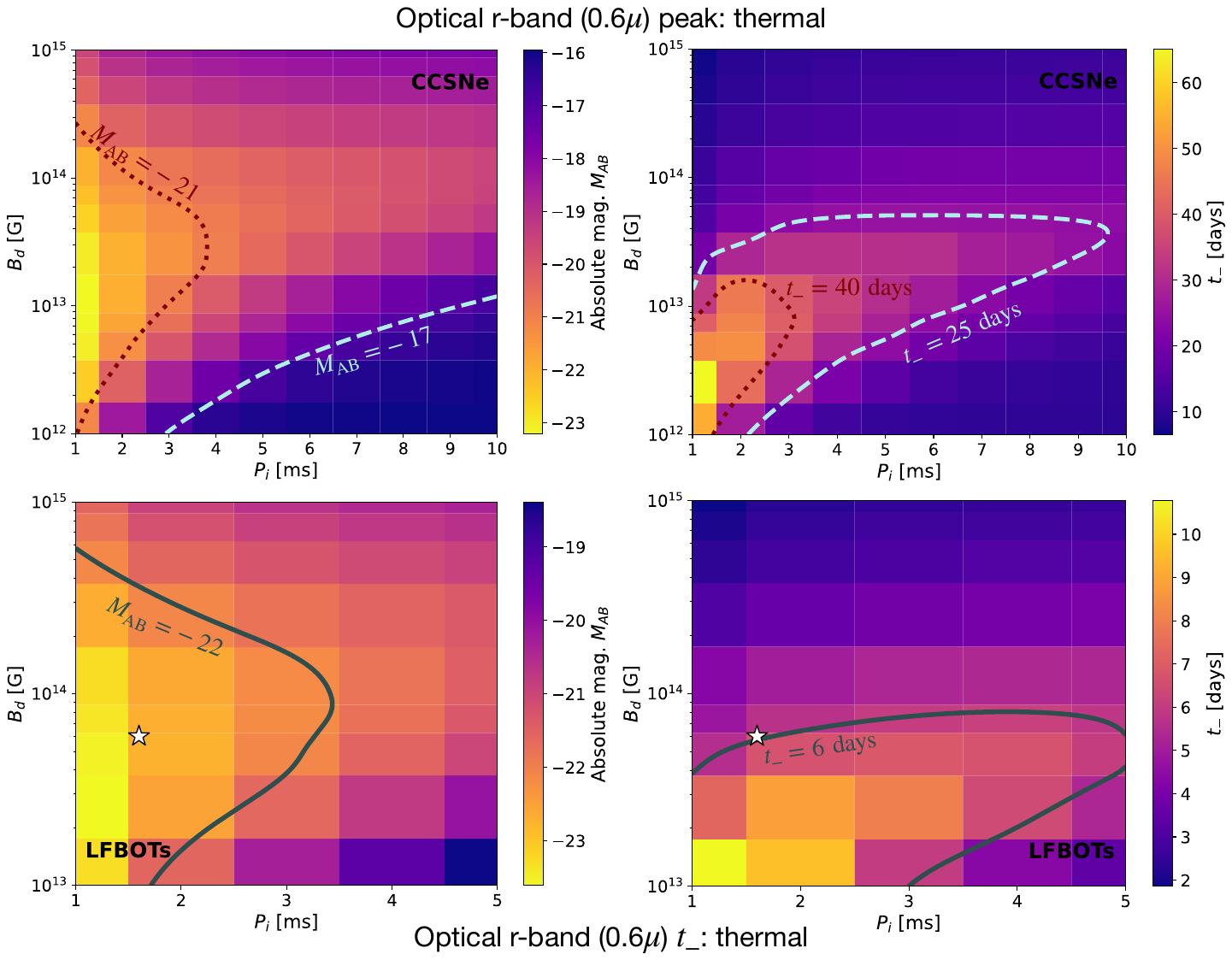}
\caption{\label{fig:mab_thalf} \emph{Left: }The absolute magnitude ($M_{\rm AB}$) of the optical thermal emission peak of r-band ($0.6\ \mu m$) for CCSNe \emph{(top)} and LFBOTs \emph{(bottom)} shown in the $B_d - P_i$ plane. For CCSNe two contours $M_{\rm AB} = -17$ (dashed light blue) and $M_{\rm AB} = -21$ (dotted maroon) denote the typical $M_{\rm AB}$ for ordinary SNe and SLSNe respectively, while for LFBOTs $M_{\rm AB} = -22$ (solid dark green) is also shown. \emph{Right: } The fall time ($t_-$: the time interval in which the peak luminosity falls to half its value) for CCSNe \emph{(top)} and LFBOTs \emph{(bottom)} shown in the same $B_d - P_i$ plane. The typical fall-time values for ordinary SNe ($t_- \sim 25$ days) and SLSNe ($t_- \sim 40$ days) are shown in the same color scheme as before, while the same for LFBOTs ($t_- \sim 6$ days) is denoted by a solid dark green line. The LFBOT parameter set ($P_i = 1.6$ ms and $B_d = 6 \times 10^{13}$ G)  (discussed in Section~\ref{sec:lfbot}) which is the dominant contributor to the diffuse flux (see Section~\ref{sec:diffuse}) is denoted using a star.
}
\end{figure*}
Having discussed LFBOTs in the previous section, in this section, we examine the potential of high-energy neutrino emission from other pulsar-powered optical transients like ordinary SNe and SLSNe along with LFBOTs, to explain the joint flux without violating constraints from optical observations. For each of them, the central engine remains the same (a pulsar remnant with $M_* = 1.42 M_\odot$ and $R_* = 12$ km) as discussed in Section~\ref{sec:model}, but the composition of the ejecta along with the ejecta mass and velocity differs. The ejecta properties for LFBOTs are already outlined in Section~\ref{sec:lfbot}. Below we discuss the relevant details about the ejecta and its composition for ordinary SNe and SLSNe, following which we finally present the results of our parameter space survey to choose the viability of the KM3NeT event.

In general, both ordinary SNe and SLSNe are associated with a fairly large ejecta mass, while the central engine for the former is much weaker than the latter. Thus for modeling ordinary SNe and SLSNe, we consider $M_{\rm ej}^{\rm SNe} = 2.23 M_\odot$ (which is $\gg M_{\rm ej}^{\rm LFBOT} = 0.1 M_\odot$), with the kinetic energy of the ejecta fixed at $\sim 5.1 \times 10^{51}$ erg similar to LFBOTs. However, due to the larger ejecta mass in this case we have the initial ejecta velocity to be $v_{\rm ej}^{\rm SNe} \approx 0.05$ c. The elemental composition of the ejecta is motivated from~\cite{Dessart:2015mga,Dessart:2016fun} (see~\citealt{Prepration:2026slsne} for more details). The SNe ejecta is mainly composed of helium, oxygen, and nickel ($1.55 M_\odot$, $0.251 M_\odot$, and $0.105 M_\odot$ respectively). The initial composition of nickel in the ejecta changes since nickel decays to cobalt which finally decays to iron over a timescale of $\mathcal{O}(100)$ days. Iron majorly alters the opacity of the ejecta. Besides, the decay of nickel also injects thermal photons in the nebula which we take into account. The ejecta opacity is chosen to be $\kappa_{\rm ej} = 0.1\ {\rm cm}^2{\rm g}^{-1}$.
\subsection{Optical properties}
To examine the optical band observables, like the peak absolute magnitude $M_{\rm AB}$ and the time during which the peak absolute magnitude decreases by $\sim 1$ defined as the fall time ($t_-$), we compute the lightcurve for the optical $r$-band ($= 0.6\ \mu m$). The peak value of the absolute AB magnitude for the thermal lightcurve in optical $r$-band is shown as a color map in the $B_d - P_i$ plane in Figure~\ref{fig:mab_thalf} \emph{(left)}. For CCSNe \emph{(top)} we survey $B_d$ between $10^{12}$ G to $10^{15}$ G, while for $P_i$ we scan between $1$ ms to $10$ ms. The color map is shown such that the brighter sources have more negative $M_{\rm AB}$ and are denoted by warmer yellow shades while the less bright sources are represented by cooler blue shades. As expected, sources with a lower initial spin period are brighter since the available spindown energy is higher to begin with. The variation with $B_d$ depends on the spindown energy $E_{\rm sd}$ and the relation between $t_{\rm sd}$ and $t_{\rm diff,0}^{\rm ej}$, where $t_{\rm diff,0}^{\rm ej}$ is the characteristic photon diffusion timescale through the ejecta. Recall that $E_{\rm sd} \propto B_d^2$ and hence increases with increase in $B_d$. However the timescale over which $E_{\rm sd}$ peaks, is given by $t_{\rm sd}\propto B_d^{-2}$, which decreases with increase in $B_d$. For a given ejecta composition, $t_{\rm diff,0}^{\rm ej}$ is decided by the ejecta opacity ($\kappa_{\rm ej}$) (see Equation~\ref{eq:photon_diff}) and remains fixed. Thus, when $t_{\rm sd} < t_{\rm diff,0}^{\rm ej}$ the peak in $E_{\rm sd}$ already occurs before the ejecta becomes transparent, leading to a lower $r-$band luminosity and vice-versa. Thus, there exists an optimal region in the $B_d - P_i$ plane where $E_{\rm sd}$ peaks close to the time when the ejecta becomes transparent, leading to brighter $r-$band luminosities.

From Figure~\ref{fig:mab_thalf} \emph{(top)}, we infer the available parameter space for ordinary SNe and SLSNe. Ordinary SNe with typical $M_{AB} \sim -17$~\citep{DES:2022vik} (shown as a dashed ligth blue contour) populate the lower right corner of the plane, with comparatively high $P_i > 3$ ms and low $B_d \sim 10^{12} - 10^{13}$ G. SLSNe are optically brighter by about two orders of magnitude and occupy the left side of the plane shown by a dark red dotted contour, $M_{AB} \sim -21$~\citep{Moriya:2024gqt}, which is typical for SLSNe. In contrast to ordinary SNe, the population of SLSNe prefers lower $P_i$ and higher $B_d$.

For LFBOTs \emph{(bottom)} we restrict our scan to $B_d$ between $10^{13} - 10^{15}$ G and $P_i$ between $1$ ms to $5$ ms. This is because for this class of transients we need the peak absolute AB magnitude in optical band to be $M_{\rm AB} \sim -22$~\citep{Perley:2018oky,Perley:2021ell} which is much easier to achieve with faster spinning pulsars as the central engine. Similar to before we show a color map of the thermal emission in the optical $r$-band where the color denotes the peak $M_{\rm AB}$. The contour corresponding to $M_{\rm AB} = -22$ is shown in solid dark green. Compared to the case of SLSNe we see much of the parameter space is suitable for $M_{\rm AB} \lesssim -22$ (that is brighter than $M_{\rm AB} \sim -21$), this has to do with the fact that  $M_{\rm ej}^{\rm FBOT} \ll M_{\rm ej}^{\rm SNe}$, reducing the attenuation and photon diffusion time through the ejecta.

A prominent feature distinguishing between each class of transient is the timescale of evolution for the lightcurve. The time interval in which the lightcurve declines to half of its peak value is defined as the fall time $t_-$ is such that $L(t_-) = L_{\rm peak}/2$. We chose $t_-$ since observationally it is easier to observe the decline of the lightcurve from its peak. The \emph{right} panels of Figure~\ref{fig:mab_thalf} show a color map where the colors denote $t_-$ of the thermal optical $r$-band lightcurve on the $B_d - P_i$ plane. For ordinary SNe typical $t_{-} \sim 25$ days, while for SLSNe $t_- \sim 40$ days~\citep{Villar:2018toe}. These contours are shown using dashed light blue and dotted dark red contours respectively as shown in the \emph{top} panel. For LFBOTs typical $t_-\sim 6$ days~\citep{Perley:2018oky}, the contour for which is shown using solid dark green line in the \emph{bottom} panel. The shorter $t_{-}$ for LFBOTs is because the photons can diffuse faster compared to both ordinary SNe and SLSNe, due to the lower ejecta mass, as the optical depth in the ejecta $\propto M_{\rm ej}$.
\subsection{Neutrino properties}
\begin{figure*}
\centering
\includegraphics[width=\textwidth]{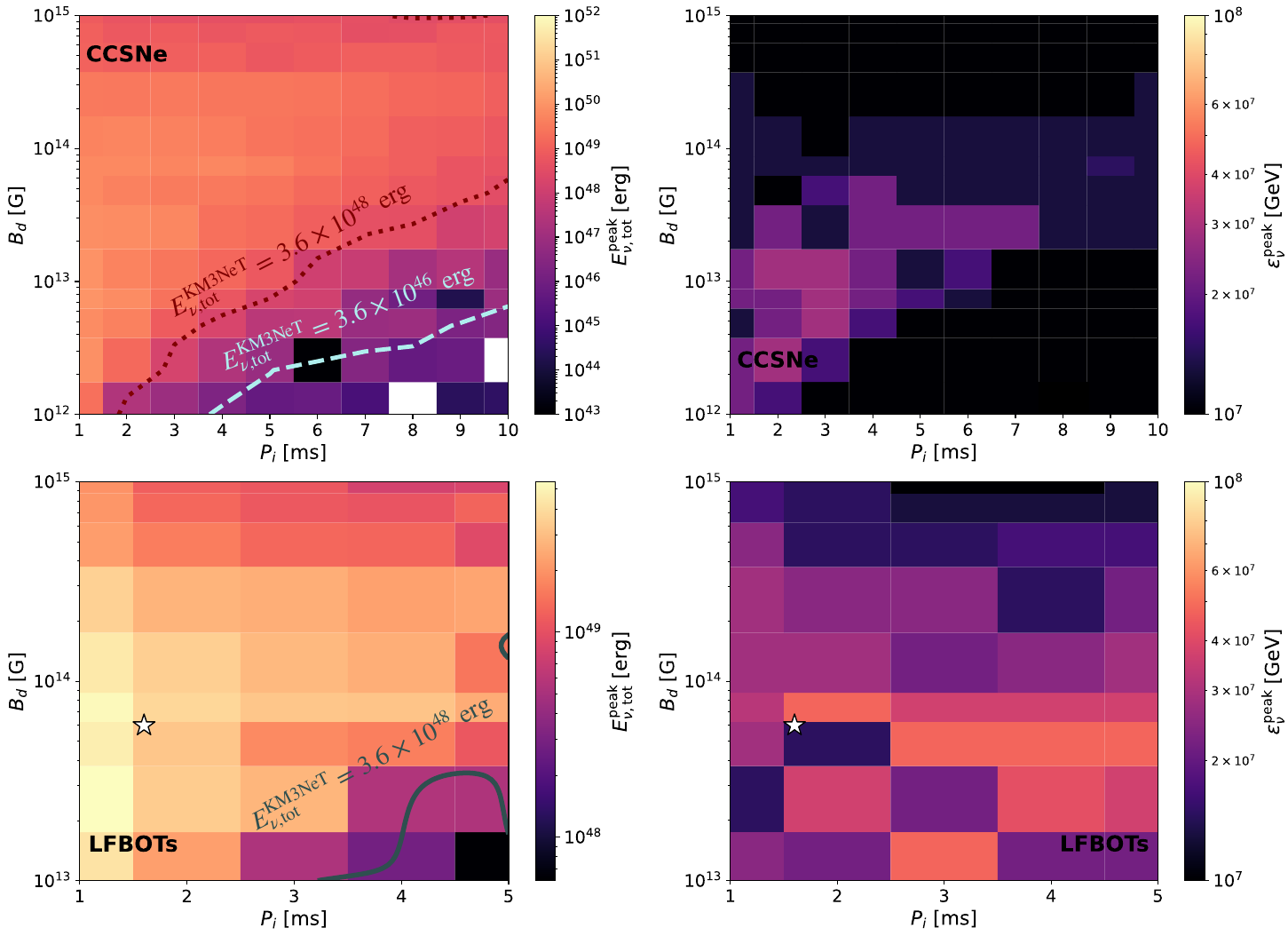}
\caption{\label{fig:fbotsenupk}The peak of total neutrino energy fluence ($E_{\nu,\rm tot}^{\rm peak}$) \emph{(left)} and the peak energy in the neutrino spectrum ($\varepsilon_\nu^{\rm peak}$) \emph{(right)} is shown in the $B_d - P_i$ plane for CCSNe \emph{(top)} and LFBOTs \emph{(bottom)}. We also show the total energy budget in neutrinos needed to match the level of the joint flux's central value of $E_\nu^2 \phi_\nu \sim 7 \times 10^{-10}{\rm GeV}{\rm cm}^{-2}{\rm s}^{-1}{\rm sr}^{-1}$ as a dashed light blue contour for ordinary SNe, a dotted maroon contour for SLSNe, and a solid dark green contour for LFBOTs assuming $\dot{\rho}_0 = 10^{-5}\ {\rm Mpc}^{-3}\rm yr^{-1}$ for ordinary SNe and $\dot{\rho}_0 = 10^{-7}\ {\rm Mpc}^{-3}\rm yr^{-1}$ for SLSNe and LFBOTs. The LFBOT parameter set ($P_i = 1.6$ ms and $B_d = 6 \times 10^{13}$ G)  (discussed in Section~\ref{sec:lfbot}) which is the dominant contributor to the diffuse flux (see Section~\ref{sec:diffuse}) is denoted using a star. Note that in the \emph{top right} panel the white squares indicate $E_{\nu,\rm tot}^{\rm peak} < 10^{43}$ erg.
}
\end{figure*}
The total energy emitted in neutrinos can be defined as $E_{\nu,\rm tot} = \int dt \int dE_\nu \left( E_\nu dN/dE_\nu\right)$. The properties of the central engine along with the radiation field in the nebula primarily govern the neutrino emission from the sources we discuss here. Although the ejecta properties are different for the various classes of transients, the characteristics associated with neutrino emission are similar. We show the peak of the total neutrino energy fluence $E_{\nu,\rm tot}^{\rm peak}$ as a color map on the $B_d - P_i$ plane in Figure~\ref{fig:fbotsenupk} \emph{(left)}. The peak energy in the neutrino spectrum, that is, the neutrino energy at which $E_{\nu,\rm tot}^{\rm peak}$ occurs, is denoted by $\varepsilon_\nu^{\rm peak}$ and is shown as a color map on the $B_d - P_i$ plane in the \emph{right} panel of Figure.~\ref{fig:fbotsenupk}.

The peak of the total energy emitted in neutrinos is roughly $\gtrsim$ a few $10^{48}$ erg for both SLSNe and LFBOTs. The neutrino energy where the peak occurs $\sim 10^7 - 10^8$ GeV. The former is decided by how much energy is injected into CR protons and what fraction of the protons cool through the photomeson channel, leading to a fraction of the total energy in protons converting to total energy in neutrinos. The latter is decided by the maximum energies to which protons are accelerated in the polar cap and TS regions and if these highly energetic protons can interact efficiently with photons to eventually produce neutrinos at the highest energies. For ordinary SNe, the total energy emitted in neutrinos is limited to $\sim$ a few $10^{46} - 10^{47}$ erg, while the neutrino energy where the peak occurs is $< 10^7$ GeV.

Next, we evaluate the single source neutrino energy fluence required to explain the joint flux. Since our hypothesis is that the neutrino event at KM3NeT resulted from a diffuse flux from an optical transient, we can use the central value of the joint flux ($\sim 7 \times 10^{-10}{\rm GeV}{\rm cm}^{-2}{\rm s}^{-1}{\rm sr}^{-1}$), to estimate the single source flux using $E_\nu^2 \big(dN_\nu/dE_\nu\big) = \left(4 \pi H_0/c\right) \big(E_\nu^2 \phi_\nu/ \left(\dot{\rho}_0 f_z \right)\big)$, where $H_0$ is the Hubble constant $H_0 = 72\ {\rm km\ s}^{-1}{\rm Mpc}^{-1}$, $\dot{\rho}_0 = \dot{\rho} (z=0)$ gives the local rate of the transients at redshift $z = 0$, and $f_z$ is the redshift factor. We choose, $f_z = 2.8$. For the local rate densities we choose, $\dot{\rho}_0^{\rm SNe} \sim 10^{-5}\ {\rm Mpc}^{-3} {\rm yr}^{-1}$, $\dot{\rho}_0^{\rm SLSNe} \sim 10^{-7}\ {\rm Mpc}^{-3} {\rm yr}^{-1}$~\citep{Moriya:2024gqt}, $\dot{\rho}_0^{\rm LFBOTs} \sim 10^{-7}\ {\rm Mpc}^{-3} {\rm yr}^{-1}$~\citep{Ho:2021fyb}. This gives the single source neutrino flux needed to match the diffuse flux level of the KM3NeT event to be $E_{\nu,\rm tot}^{\rm KM3NeT} \sim 3.6 \times 10^{46}\ {\rm erg}$ for ordinary SNe and $E_{\nu,\rm tot}^{\rm KM3NeT} \sim 3.6 \times 10^{48}\ {\rm erg}$ for SLSNe and LFBOTs. The contours of $E_{\nu,\rm tot}^{\rm KM3NeT}$ are shown as dashed light blue contour, dotted dark red contour, and solid dark green contour corresponding to ordinary SNe, SLSNe, and LFBOTs respectively, in Figure~\ref{fig:fbotsenupk} \emph{(left)}.

Figures~\ref{fig:mab_thalf} and~\ref{fig:fbotsenupk} together help us identify a plausible class of optical transients that satisfy two crucial criteria: (a) produce high energy neutrinos similar to the neutrino energy for the joint flux ($\mathcal{O}(100)$ PeV) and (b) the total energy emitted in neutrinos and the rate of such transients is consistent with the diffuse emission estimated from the joint flux. For ordinary SNe, the requirement of matching the diffuse limit from the joint flux ($E_{\nu,\rm tot}^{\rm KM3NeT} \sim 3.6 \times 10^{46}\ {\rm erg}$) still allows a part of the parameter space not excluded $M_{\rm AB} = -17$. However, $\varepsilon_\nu^{\rm peak} < 10^7$ GeV, excluding them as possible sources contributing to the diffuse flux explaining the KM3NeT event. This is to be expected since in ordinary SNe owing to low $B_d$ and high $P_i$, the available spindown energy is lower, leading to a less powerful central engine than SLSNe and LFBOTs.

Both SLSNe and LFBOTs are good candidates to explain the KM3NeT event since they have a powerful central engine that have a comparatively large reservoir of spindown energy from the pulsar. This is attributed to the higher $B_d$ and lower $P_i$ for these systems. The higher spindown energy also makes these systems brighter in the electromagnetic channel and hence typical optical luminosities from these systems are $M_{AB}\sim - 21$ (SLSNe) or $M_{AB}\sim - 22$ (LFBOTs). This however also makes these systems rarer and hence they have a reduced fiducial rate $\dot{\rho}_0 \sim 10^{-7}\ {\rm Mpc}^{-3} {\rm yr}^{-1}$. The required diffuse neutrino flux limit is easily satisified by both SLSNe and LFBOTs. However, LFBOTs, owing to their smaller ejecta masses and the powerful central engine, can provide a larger population of objects that satisfy the conditions on $\varepsilon_\nu^{\rm peak}$. This is crucial for satisfying the diffuse limits, which involve contributions from a population of this transient class with different parameters. Although individual SLSNe can have $\varepsilon_\nu^{\rm peak}$ comparable to individual LFBOTs for similar central engine parameters ($B_d$, $P_i$), the diffuse neutrino spectrum from the LFBOT population peaks at slightly higher energies due to their systematically higher population-averaged $\varepsilon_\nu^{\rm peak}$.
\section{Diffuse contribution: explaining the KM3N\MakeLowercase{e}T event}
\label{sec:diffuse}
\begin{figure*}[ht!]
\centering
\includegraphics[width=0.99\textwidth]{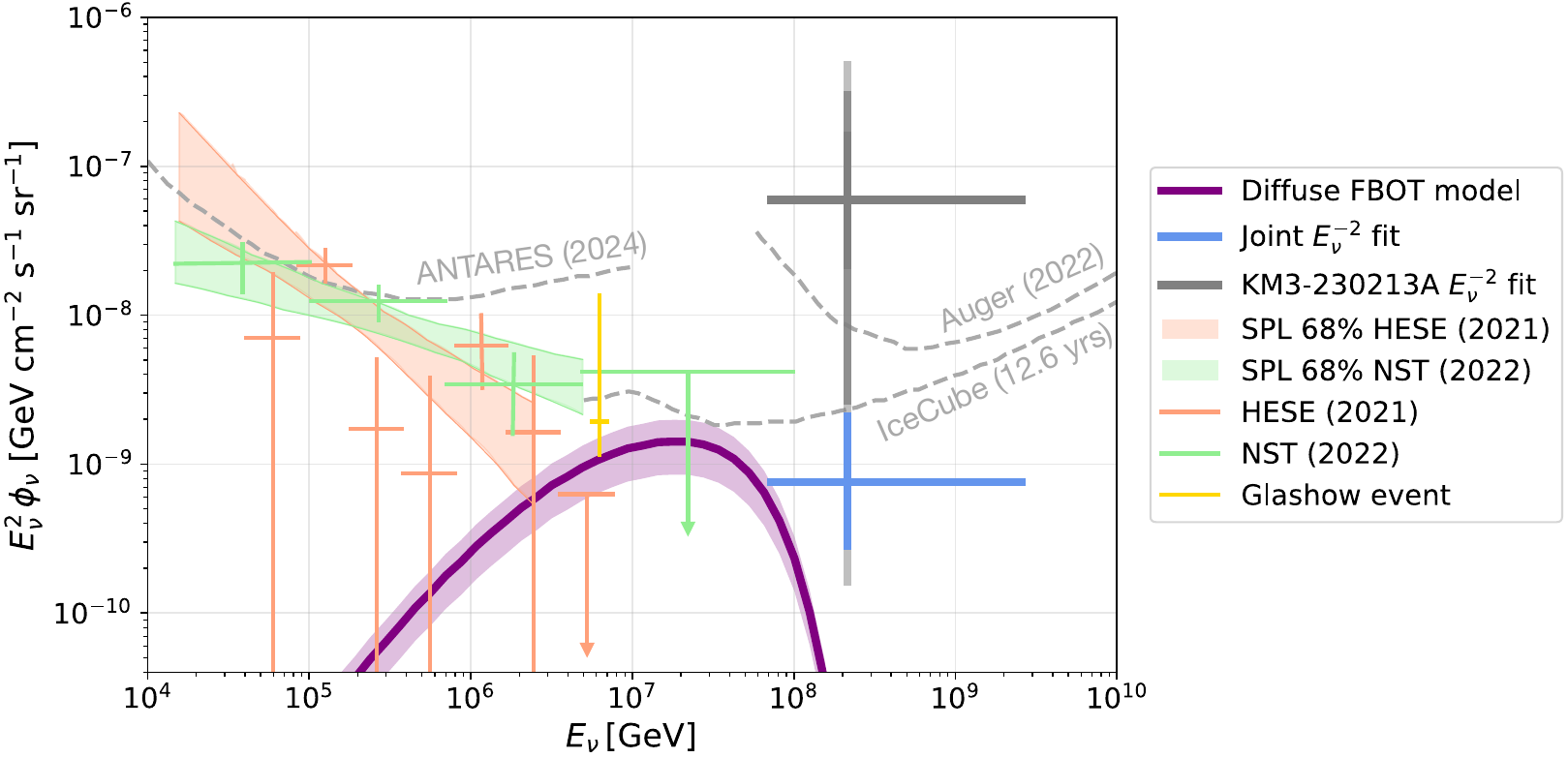}
\caption{\label{fig:res}The diffuse high energy neutrino spectrum from a population of LFBOTs (see Equation~\ref{eq:diffuse} and Section~\ref{appsec:diffuse_comp}). The shaded purple band shows the 40\% uncertainty in the diffuse flux. The light blue cross denotes the \emph{joint flux} obtained from~\cite{KM3NeT:2025ccp} by including IceCube Extreme High-energy (IC-EHE)~\citep{IceCube:2018fhm} and Auger non-observations along with the updated IceCube ~\citep{IceCubeCollaborationSS:2025jbi} and differential upper limits from Auger~\citep{PierreAuger:2023pjg}. The diffuse upper limits from ANTARES at 95\% C.L.~\citep{ANTARES:2024ihw}, Auger at 90\% C.L.~\citep{PierreAuger:2023pjg}, and IC-EHE at 90\% C.L.~\citep{IceCube:2018fhm} are shown as dashed gray lines. The light orange and light green shaded regions represent the 68\% C.L. contours of the IceCube single power-law (SPL) fits to the Northern Sky Tracks (NST)~\citep{Abbasi:2021qfz} and High Energy Starting Events (HESE)~\citep{IceCube:2020wum} datasets. The segmented fits from the same datasets are shown with light orange and light green crosses respectively. Additionally, the Glashow resonance event at IceCube~\citep{IceCube:2021rpz} is shown in yellow. The equivalent flux for KM3-230213A~\citep{KM3NeT:2025npi} is shown using gray lines along with the $1 \sigma$, $2 \sigma$, and $3 \sigma$ uncertainty bands in darker shades of gray.
}
\end{figure*}
Having established LFBOTs as the primary candidate to explain the KM3NeT event as a diffuse contribution from the class of pulsar-powered optical transients (ordinary SNe and SLSNe), in this section we deal with evaluating the diffuse contribution of high energy neutrinos from LFBOTs. In particular, we try to answer the primary question that we had posed in the Introduction (Section~\ref{sec:intro}), that is, can a diffuse neutrino contribution from LFBOTs explain the joint flux?

We outline the formalism and provide details regarding the computation of the diffuse neutrino flux in Appendix~\ref{appsec:diffuse_comp}. Our main result obtained using Equation~\eqref{eq:diffuse}, is shown in Figure~\ref{fig:res}. The solid purple line denotes the resulting diffuse neutrino flux from a population of LFBOTs, while the light purple shaded band denotes a 40\% uncertainty. The most important takeaway from Figure~\ref{fig:res} is that indeed a diffuse flux from LFBOTs can serve as a plausible explanation to the observed KM3-230213A event based on the joint flux results, as can be seen from the overlap between portions of the blue cross and the purple band.
\section{Summary and discussion}
\label{sec:disc}
We investigated the role of pulsar-powered optical transients in explaining the recently observed $\sim 220$ PeV neutrino event at KM3NeT, KM3-230213A. We examined the potential of three distinct class of transients resulting from core-collapse (ordinary SNe, SLSNe, and LFBOTs) to produce high-energy neutrino signatures. Using the KM3NeT event, along with non-observations from the IceCube and Pierre Auger observatories, a joint fit can be obtained for the diffuse UHE neutrino flux, reconciling KM3-230213A to the limits from IceCube and Auger. Our aim in this work was to evaluate which population of pulsar-powered optical transients can match the flux level and energy of the joint flux.

The two main parameters associated with the central engine of all the optical transients studied in this work are the initial spin period ($P_i$) and the dipolar magnetic field strength ($B_d$) of the newly born pulsar. We consider that the central engines for ordinary SNe and SLSNe are embedded in a heavier ejecta, while LFBOTs have a smaller ejecta mass. We surveyed the $B_d-P_i$ plane for the transients (see Section~\ref{sec:survey}) in terms of the peak $M_{\rm AB}$ magnitude and the fall time of the r-band lightcurves (Figure~\ref{fig:mab_thalf}), the peak of the total energy fluence in neutrinos and the peak energy in the neutrino spectrum (Figure~\ref{fig:fbotsenupk}). The former serve as characteristics to determine regions of the parameter space suitable for the transients, while the latter allow us to determine which class of transients can possibly explain the joint flux through a diffuse contribution. We show our main result in Figure~\ref{fig:res}, which highlights that a diffuse high-energy neutrino flux from a population of LFBOTs (discussed in Section~\ref{sec:lfbot}, Figures~\ref{fig:lbol_temp_lc} and~\ref{fig:nu}) can explain the joint flux.

The advent of EM observatories like,  the Vera C. Rubin Observatory Legacy Survey of Space and Time (LSST)~\citep{LSST:2008ijt,Blum:2022dxi}, greatly improve the available data for optical transients. Additionally, several current and upcoming UHE neutrino observatories~\citep{Kotera:2025jca} like RNO-G~\citep{RNO-G:2020rmc}, GRAND~\citep{GRAND:2018iaj}, BEACON~\citep{BEACON:2025qcq}, PUEO~\citep{PUEO:2020bnn}, IceCube-Gen2 Radio~\citep{IceCube-Gen2:2020qha}, TAMBO~\citep{TAMBO:2025jio}, Trinity~\citep{Otte:2025dld}, and POEMMA or SPB-2~\citep{POEMMA:2020ykm} will contribute to better statistics associated with UHE neutrinos. Thus, the question whether KM3-230213A originated as a result of a diffuse contribution from optical pulsar-powered transients can be concretely answered using stacking searches. In fact, due to the associated modeling uncertainties, besides LFBOTs, we consistently survey the parameter space for both ordinary SNe and SLSNe which can also contribute to the diffuse UHE neutrino flux. Our current work highlights that possibility and serves as a step in that direction.
\vspace{-2em}
\begin{acknowledgements}
We thank Conor Omand for providing us with useful suggestions.
M.\,M. wishes to thank the organizers of WHEPP 2025 where this idea initiated and where the majority of the work was completed.
M.\,M. acknowledges support from the FermiForward Discovery Group, LLC under Contract No. 89243024CSC000002 with the U.S. Department of Energy, Office of Science, Office of High Energy Physics. S.S.K. acknowledges the support by KAKENHI No. 22K14028, No. 21H04487, No. 23H04899, and the Tohoku Initiative for Fostering Global Researchers for Interdisciplinary Sciences (TI-FRIS) of MEXT’s Strategic Professional Development Program for Young Researchers.
\end{acknowledgements}
\bibliography{refs}{}

\begin{thebibliography}{}
\expandafter\ifx\csname natexlab\endcsname\relax\def\natexlab#1{#1}\fi
\providecommand{\url}[1]{\href{#1}{#1}}
\providecommand{\dodoi}[1]{doi:~\href{http://doi.org/#1}{\nolinkurl{#1}}}
\providecommand{\doeprint}[1]{\href{http://ascl.net/#1}{\nolinkurl{http://ascl.net/#1}}}
\providecommand{\doarXiv}[1]{\href{https://arxiv.org/abs/#1}{\nolinkurl{https://arxiv.org/abs/#1}}}

% type= article
\bibitem[{M.~G. Aartsen {et~al.}(2018)Aartsen {et~al.}}]{IceCube:2018fhm}
Aartsen, M.~G., {et~al.} 2018, \bibinfo{title}{{Differential limit on the extremely-high-energy cosmic neutrino flux in the presence of astrophysical background from nine years of IceCube data},} Phys. Rev. D, 98, 062003, \dodoi{10.1103/PhysRevD.98.062003}

% type= article
\bibitem[{M.~G. Aartsen {et~al.}(2021{\natexlab{a}})Aartsen {et~al.}}]{IceCube:2021rpz}
Aartsen, M.~G., {et~al.} 2021{\natexlab{a}}, \bibinfo{title}{{Detection of a particle shower at the Glashow resonance with IceCube},} Nature, 591, 220, \dodoi{10.1038/s41586-021-03256-1}

% type= article
\bibitem[{M.~G. Aartsen {et~al.}(2021{\natexlab{b}})Aartsen {et~al.}}]{IceCube-Gen2:2020qha}
Aartsen, M.~G., {et~al.} 2021{\natexlab{b}}, \bibinfo{title}{{IceCube-Gen2: the window to the extreme Universe},} J. Phys. G, 48, 060501, \dodoi{10.1088/1361-6471/abbd48}

% type= article
\bibitem[{Q. Abarr {et~al.}(2021)Abarr {et~al.}}]{PUEO:2020bnn}
Abarr, Q., {et~al.} 2021, \bibinfo{title}{{The Payload for Ultrahigh Energy Observations (PUEO): a white paper},} JINST, 16, P08035, \dodoi{10.1088/1748-0221/16/08/P08035}

% type= article
\bibitem[{R. Abbasi {et~al.}(2021)Abbasi {et~al.}}]{IceCube:2020wum}
Abbasi, R., {et~al.} 2021, \bibinfo{title}{{The IceCube high-energy starting event sample: Description and flux characterization with 7.5 years of data},} Phys. Rev. D, 104, 022002, \dodoi{10.1103/PhysRevD.104.022002}

% type= article
\bibitem[{R. Abbasi {et~al.}(2022)Abbasi {et~al.}}]{Abbasi:2021qfz}
Abbasi, R., {et~al.} 2022, \bibinfo{title}{{Improved Characterization of the Astrophysical Muon{\textendash}neutrino Flux with 9.5 Years of IceCube Data},} Astrophys. J., 928, 50, \dodoi{10.3847/1538-4357/ac4d29}

% type= article
\bibitem[{R. Abbasi {et~al.}(2025)Abbasi {et~al.}}]{IceCubeCollaborationSS:2025jbi}
Abbasi, R., {et~al.} 2025, \bibinfo{title}{{Search for Extremely-High-Energy Neutrinos and First Constraints on the Ultrahigh-Energy Cosmic-Ray Proton Fraction with IceCube},} Phys. Rev. Lett., 135, 031001, \dodoi{10.1103/PhysRevLett.135.031001}

% type= article
\bibitem[{A. Abdul~Halim {et~al.}(2023)Abdul~Halim {et~al.}}]{PierreAuger:2023pjg}
Abdul~Halim, A., {et~al.} 2023, \bibinfo{title}{{Latest results from the searches for ultra-high-energy photons and neutrinos at the Pierre Auger Observatory},} PoS, ICRC2023, 1488, \dodoi{10.22323/1.444.1488}

% type= article
\bibitem[{O. Adriani {et~al.}(2025{\natexlab{a}})Adriani {et~al.}}]{KM3NeT:2025ccp}
Adriani, O., {et~al.} 2025{\natexlab{a}}, \bibinfo{title}{{Ultrahigh-Energy Event KM3-230213A within the Global Neutrino Landscape},} Phys. Rev. X, 15, 031016, \dodoi{10.1103/yypk-zmb8}

% type= article
\bibitem[{O. Adriani {et~al.}(2025{\natexlab{b}})Adriani {et~al.}}]{KM3NeT:2025bxl}
Adriani, O., {et~al.} 2025{\natexlab{b}}, \bibinfo{title}{{Characterizing Candidate Blazar Counterparts of the Ultra-High-Energy Event KM3-230213A},} \doarXiv{2502.08484}

% type= article
\bibitem[{O. Adriani {et~al.}(2025{\natexlab{c}})Adriani {et~al.}}]{KM3NeT:2025lly}
Adriani, O., {et~al.} 2025{\natexlab{c}}, \bibinfo{title}{{Blazars as a Potential Origin of the KM3-230213A Event},} \doarXiv{2511.13886}

% type= article
\bibitem[{O. Adriani {et~al.}(2025{\natexlab{d}})Adriani {et~al.}}]{KM3NeT:2025aps}
Adriani, O., {et~al.} 2025{\natexlab{d}}, \bibinfo{title}{{On the Potential Galactic Origin of the Ultra-High-Energy Event KM3-230213A},} \doarXiv{2502.08387}

% type= article
\bibitem[{J.~A. Aguilar {et~al.}(2021)Aguilar {et~al.}}]{RNO-G:2020rmc}
Aguilar, J.~A., {et~al.} 2021, \bibinfo{title}{{Design and Sensitivity of the Radio Neutrino Observatory in Greenland (RNO-G)},} JINST, 16, P03025, \dodoi{10.1088/1748-0221/16/03/P03025}

% type= article
\bibitem[{S. Aiello {et~al.}(2025)Aiello {et~al.}}]{KM3NeT:2025npi}
Aiello, S., {et~al.} 2025, \bibinfo{title}{{Observation of an ultra-high-energy cosmic neutrino with KM3NeT},} Nature, 638, 376, \dodoi{10.1038/s41586-024-08543-1}

% type= article
\bibitem[{L.~F.~T. Airoldi {et~al.}(2025)Airoldi, Alves, Perez-Gonzalez, Salla, \& Funchal}]{Airoldi:2025opo}
Airoldi, L. F.~T., Alves, G. F.~S., Perez-Gonzalez, Y.~F., Salla, G.~M., \& Funchal, R.~Z. 2025, \bibinfo{title}{{Could a Primordial Black Hole Explosion Explain the extremely high-energy KM3NeT neutrino Event?},} \doarXiv{2505.24666}

% type= article
\bibitem[{A. Albert {et~al.}(2024)Albert {et~al.}}]{ANTARES:2024ihw}
Albert, A., {et~al.} 2024, \bibinfo{title}{{Constraints on the energy spectrum of the diffuse cosmic neutrino flux from the ANTARES neutrino telescope},} JCAP, 08, 038, \dodoi{10.1088/1475-7516/2024/08/038}

% type= article
\bibitem[{R. Aloisio {et~al.}(2025)Aloisio, Ambrosone, \& Evoli}]{Aloisio:2025nts}
Aloisio, R., Ambrosone, A., \& Evoli, C. 2025, \bibinfo{title}{{Constraining Super-Heavy Dark Matter with the KM3-230213A Neutrino Event},} \doarXiv{2508.08779}

% type= article
\bibitem[{J. {\'A}lvarez-Mu{\~n}iz {et~al.}(2020){\'A}lvarez-Mu{\~n}iz {et~al.}}]{GRAND:2018iaj}
{\'A}lvarez-Mu{\~n}iz, J., {et~al.} 2020, \bibinfo{title}{{The Giant Radio Array for Neutrino Detection (GRAND): Science and Design},} Sci. China Phys. Mech. Astron., 63, 219501, \dodoi{10.1007/s11433-018-9385-7}

% type= article
\bibitem[{C.~A. Arg{\"u}elles {et~al.}(2025)Arg{\"u}elles {et~al.}}]{TAMBO:2025jio}
Arg{\"u}elles, C.~A., {et~al.} 2025, \bibinfo{title}{{TAMBO: A Deep-Valley Neutrino Observatory},} \doarXiv{2507.08070}

% type= article
\bibitem[{J. Arons(2003)Arons}]{Arons:2002yj}
Arons, J. 2003, \bibinfo{title}{{Magnetars in the metagalaxy: an origin for ultrahigh-energy cosmic rays in the nearby universe},} Astrophys. J., 589, 871, \dodoi{10.1086/374776}

% type= article
\bibitem[{P. Blasi {et~al.}(2000)Blasi, Epstein, \& Olinto}]{Blasi:2000xm}
Blasi, P., Epstein, R.~I., \& Olinto, A.~V. 2000, \bibinfo{title}{{Ultrahigh-energy cosmic rays from young neutron star winds},} Astrophys. J. Lett., 533, L123, \dodoi{10.1086/312626}

% type= inproceedings
\bibitem[{B. Blum {et~al.}(2022)Blum {et~al.}}]{Blum:2022dxi}
Blum, B., {et~al.} 2022, \bibinfo{title}{{Snowmass2021 Cosmic Frontier White Paper: Rubin Observatory after LSST},} in {Snowmass 2021}.
\newblock \doarXiv{2203.07220}

% type= article
\bibitem[{A. Boccia \& F. Iocco(2025)Boccia \& Iocco}]{Boccia:2025hpm}
Boccia, A., \& Iocco, F. 2025, \bibinfo{title}{{Could the KM3{\textendash}230213A event be caused by an evaporating primordial black hole?},} Phys. Rev. D, 112, 063045, \dodoi{10.1103/qxcj-fpwn}

% type= article
\bibitem[{D. Borah {et~al.}(2025)Borah, Das, Okada, \& Sarmah}]{Borah:2025igh}
Borah, D., Das, N., Okada, N., \& Sarmah, P. 2025, \bibinfo{title}{{Possible origin of the KM3-230213A neutrino event from dark matter decay},} Phys. Rev. D, 111, 123022, \dodoi{10.1103/shsw-mct6}

% type= article
\bibitem[{S. Boxi {et~al.}(2026)Boxi, Das, \& Gupta}]{Boxi:2025ony}
Boxi, S., Das, S., \& Gupta, N. 2026, \bibinfo{title}{{Cosmogenic Origin of KM3-230213A: Delayed Gamma-Ray Emission from a Cosmic-Ray Transient},} Astrophys. J. Lett., 997, L3, \dodoi{10.3847/2041-8213/ae3082}

% type= article
\bibitem[{G.~F. de~Clairfontaine {et~al.}(2025)de~Clairfontaine, Perucho, \& Mart{\'\i}}]{deClairfontaine:2025gei}
de~Clairfontaine, G.~F., Perucho, M., \& Mart{\'\i}, J.~M. 2025, \bibinfo{title}{{Jet-red giant interactions as a source of extragalactic neutrinos: Insights from KM3-230213A},} \doarXiv{2511.01729}

% type= article
\bibitem[{L. Dessart {et~al.}(2015)Dessart, Hillier, Woosley, Livne, Waldman, Yoon, \& Langer}]{Dessart:2015mga}
Dessart, L., Hillier, D.~J., Woosley, S., {et~al.} 2015, \bibinfo{title}{{Radiative-transfer models for supernovae IIb/Ib/Ic from binary-star progenitors},} Mon. Not. Roy. Astron. Soc., 453, 2189, \dodoi{10.1093/mnras/stv1747}

% type= article
\bibitem[{L. Dessart {et~al.}(2016)Dessart, Hillier, Woosley, Livne, Waldman, Yoon, \& Langer}]{Dessart:2016fun}
Dessart, L., Hillier, D.~J., Woosley, S., {et~al.} 2016, \bibinfo{title}{{Inferring supernova IIb/Ib/Ic ejecta properties from light curves and spectra: Correlations from radiative-transfer models},} Mon. Not. Roy. Astron. Soc., 458, 1618, \dodoi{10.1093/mnras/stw418}

% type= article
\bibitem[{P.~S.~B. Dev {et~al.}(2025)Dev, Dutta, Karthikeyan, Maitra, Strigari, \& Verma}]{Dev:2025czz}
Dev, P. S.~B., Dutta, B., Karthikeyan, A., {et~al.} 2025, \bibinfo{title}{{`Dark' Matter Effect as a Novel Solution to the KM3-230213A Puzzle},} \doarXiv{2505.22754}

% type= article
\bibitem[{T. Dzhatdoev(2025)Dzhatdoev}]{Dzhatdoev:2025sdi}
Dzhatdoev, T. 2025, \bibinfo{title}{{The blazar PKS 0605-085 as the origin of the KM3-230213A neutrino event},} PoS, ICRC2025, 1032, \dodoi{10.22323/1.501.1032}

% type= article
\bibitem[{K. Fang(2015)Fang}]{Fang:2014qva}
Fang, K. 2015, \bibinfo{title}{{High-Energy Neutrino Signatures of Newborn Pulsars In the Local Universe},} JCAP, 06, 004, \dodoi{10.1088/1475-7516/2015/06/004}

% type= article
\bibitem[{K. Fang {et~al.}(2014)Fang, Kotera, Murase, \& Olinto}]{Fang:2013vla}
Fang, K., Kotera, K., Murase, K., \& Olinto, A.~V. 2014, \bibinfo{title}{{Testing the Newborn Pulsar Origin of Ultrahigh Energy Cosmic Rays with EeV Neutrinos},} Phys. Rev. D, 90, 103005, \dodoi{10.1103/PhysRevD.90.103005}

% type= article
\bibitem[{K. Fang {et~al.}(2016)Fang, Kotera, Murase, \& Olinto}]{Fang:2015xhg}
Fang, K., Kotera, K., Murase, K., \& Olinto, A.~V. 2016, \bibinfo{title}{{IceCube Constraints on Fast-Spinning Pulsars as High-Energy Neutrino Sources},} JCAP, 04, 010, \dodoi{10.1088/1475-7516/2016/04/010}

% type= article
\bibitem[{K. Fang {et~al.}(2019)Fang, Metzger, Murase, Bartos, \& Kotera}]{Fang:2018hjp}
Fang, K., Metzger, B.~D., Murase, K., Bartos, I., \& Kotera, K. 2019, \bibinfo{title}{{Multimessenger Implications of AT2018cow: High-energy Cosmic-Ray and Neutrino Emissions from Magnetar-powered Superluminous Transients},} Astrophys. J., 878, 34, \dodoi{10.3847/1538-4357/ab1b72}

% type= article
\bibitem[{P. Goldreich \& W.~H. Julian(1969)Goldreich \& Julian}]{Goldreich:1969sb}
Goldreich, P., \& Julian, W.~H. 1969, \bibinfo{title}{{Pulsar electrodynamics},} Astrophys. J., 157, 869, \dodoi{10.1086/150119}

% type= article
\bibitem[{M. Grayling {et~al.}(2023)Grayling {et~al.}}]{DES:2022vik}
Grayling, M., {et~al.} 2023, \bibinfo{title}{{Core-collapse supernovae in the Dark Energy Survey: luminosity functions and host galaxy demographics},} Mon. Not. Roy. Astron. Soc., 520, 684, \dodoi{10.1093/mnras/stad056}

% type= article
\bibitem[{E. Guarini {et~al.}(2022)Guarini, Tamborra, \& Margutti}]{Guarini:2022uyp}
Guarini, E., Tamborra, I., \& Margutti, R. 2022, \bibinfo{title}{{Neutrino Emission from Luminous Fast Blue Optical Transients},} Astrophys. J., 935, 157, \dodoi{10.3847/1538-4357/ac7fa0}

% type= article
\bibitem[{D. Harari {et~al.}(2014)Harari, Mollerach, \& Roulet}]{Harari:2013pea}
Harari, D., Mollerach, S., \& Roulet, E. 2014, \bibinfo{title}{{Anisotropies of ultrahigh energy cosmic rays diffusing from extragalactic sources},} Phys. Rev. D, 89, 123001, \dodoi{10.1103/PhysRevD.89.123001}

% type= article
\bibitem[{A.~Y.~Q. Ho {et~al.}(2023)Ho {et~al.}}]{Ho:2021fyb}
Ho, A. Y.~Q., {et~al.} 2023, \bibinfo{title}{{A Search for Extragalactic Fast Blue Optical Transients in ZTF and the Rate of AT2018cow-like Transients},} Astrophys. J., 949, 120, \dodoi{10.3847/1538-4357/acc533}

% type= article
\bibitem[{S. Horiuchi {et~al.}(2008)Horiuchi, Suwa, Takami, Ando, \& Sato}]{Horiuchi:2008zc}
Horiuchi, S., Suwa, Y., Takami, H., Ando, S., \& Sato, K. 2008, \bibinfo{title}{{Nonthermal neutrinos from supernovae leaving a magnetar},} Mon. Not. Roy. Astron. Soc., 391, 1893, \dodoi{10.1111/j.1365-2966.2008.14000.x}

% type= article
\bibitem[{{\v{Z}}. Ivezi{\'c} {et~al.}(2019)Ivezi{\'c} {et~al.}}]{LSST:2008ijt}
Ivezi{\'c}, {\v{Z}}., {et~al.} 2019, \bibinfo{title}{{LSST: from Science Drivers to Reference Design and Anticipated Data Products},} Astrophys. J., 873, 111, \dodoi{10.3847/1538-4357/ab042c}

% type= article
\bibitem[{K. Kashiyama {et~al.}(2016)Kashiyama, Murase, Bartos, Kiuchi, \& Margutti}]{Kashiyama:2015eua}
Kashiyama, K., Murase, K., Bartos, I., Kiuchi, K., \& Margutti, R. 2016, \bibinfo{title}{{Multi-Messenger Tests for Fast-Spinning Newborn Pulsars Embedded in Stripped-Envelope Supernovae},} Astrophys. J., 818, 94, \dodoi{10.3847/0004-637X/818/1/94}

% type= article
\bibitem[{S. Khan {et~al.}(2025)Khan, Kim, \& Ko}]{Khan:2025gxs}
Khan, S., Kim, J., \& Ko, P. 2025, \bibinfo{title}{{Linking the KM3-230213A neutrino event to dark matter decay and gravitational wave signals},} JCAP, 11, 033, \dodoi{10.1088/1475-7516/2025/11/033}

% type= article
\bibitem[{K. Kohri {et~al.}(2025)Kohri, Paul, \& Sahu}]{Kohri:2025bsn}
Kohri, K., Paul, P.~K., \& Sahu, N. 2025, \bibinfo{title}{{Super heavy dark matter origin of the PeV neutrino event: KM3-230213A},} Phys. Rev. D, 112, L031703, \dodoi{10.1103/vvqq-1z2t}

% type= article
\bibitem[{K. Kotera {et~al.}(2025)Kotera, Mukhopadhyay, Alves~Batista, Fox, Martineau-Huynh, Murase, Wissel, \& Zeolla}]{Kotera:2025jca}
Kotera, K., Mukhopadhyay, M., Alves~Batista, R., {et~al.} 2025, \bibinfo{title}{{Observational strategies for ultrahigh-energy neutrinos: the importance of deep sensitivity for detection and astronomy},} \doarXiv{2504.08973}

% type= article
\bibitem[{M.~Y. Kuznetsov {et~al.}(2025)Kuznetsov, Petrov, \& Savchenko}]{Kuznetsov:2025ehl}
Kuznetsov, M.~Y., Petrov, N.~A., \& Savchenko, Y.~S. 2025, \bibinfo{title}{{Ultra-high energy event KM3-230213A as a cosmogenic neutrino in light of minimal UHECR flux models},} \doarXiv{2509.09590}

% type= article
\bibitem[{P. Madau \& M. Dickinson(2014)Madau \& Dickinson}]{Madau:2014bja}
Madau, P., \& Dickinson, M. 2014, \bibinfo{title}{{Cosmic Star Formation History},} Ann. Rev. Astron. Astrophys., 52, 415, \dodoi{10.1146/annurev-astro-081811-125615}

% type= article
\bibitem[{T.~J. Moriya(2024)Moriya}]{Moriya:2024gqt}
Moriya, T.~J. 2024, \bibinfo{title}{{Superluminous supernovae},} \doarXiv{2407.12302}

% type= article
\bibitem[{M. Mukhopadhyay \& S.~S. Kimura(2025)Mukhopadhyay \& Kimura}]{Mukhopadhyay:2025tvz}
Mukhopadhyay, M., \& Kimura, S.~S. 2025, \bibinfo{title}{{Electromagnetic Signatures from Pulsar Remnants of Binary Neutron Star Mergers: Prospects for Unique Identification Using Multiwavelength Signatures},} Astrophys. J. Lett., 989, L41, \dodoi{10.3847/2041-8213/adf285}

% type= article
\bibitem[{M. Mukhopadhyay {et~al.}(2025)Mukhopadhyay, Kimura, \& Metzger}]{Mukhopadhyay:2024ehs}
Mukhopadhyay, M., Kimura, S.~S., \& Metzger, B.~D. 2025, \bibinfo{title}{{High-energy neutrino signatures from pulsar remnants of binary neutron-star mergers: coincident detection prospects with gravitational waves},} Astrophys. J., 987, 2, \dodoi{10.3847/1538-4357/adc913}

% type= article
\bibitem[{M. Mukohpadhyay {et~al.}(2026)Mukohpadhyay, Kimura, Vurm, Metzger, \& Kashiyama}]{Prepration:2026slsne}
Mukohpadhyay, M., Kimura, S.~S., Vurm, I., Metzger, B.~D., \& Kashiyama, K. 2026, \bibinfo{title}{{Multi-messenger emissions from Super-Luminous Supernovae (SLSNe)},} (in prep)

% type= article
\bibitem[{K. Murase {et~al.}(2009)Murase, Meszaros, \& Zhang}]{Murase:2009pg}
Murase, K., Meszaros, P., \& Zhang, B. 2009, \bibinfo{title}{{Probing the birth of fast rotating magnetars through high-energy neutrinos},} Phys. Rev. D, 79, 103001, \dodoi{10.1103/PhysRevD.79.103001}

% type= article
\bibitem[{K. Murase {et~al.}(2025)Murase, Narita, \& Yin}]{Murase:2025uwv}
Murase, K., Narita, Y., \& Yin, W. 2025, \bibinfo{title}{{Superheavy dark matter from the natural inflation in light of the highest-energy astroparticle events},} JCAP, 10, 109, \dodoi{10.1088/1475-7516/2025/10/109}

% type= article
\bibitem[{Y. Narita \& W. Yin(2025)Narita \& Yin}]{Narita:2025udw}
Narita, Y., \& Yin, W. 2025, \bibinfo{title}{{Explaining the KM3-230213A Detection without Gamma-Ray Emission: Cosmic-Ray Dark Radiation},} \doarXiv{2503.07776}

% type= article
\bibitem[{A. Neronov {et~al.}(2025)Neronov, Oikonomou, \& Semikoz}]{Neronov:2025jfj}
Neronov, A., Oikonomou, F., \& Semikoz, D. 2025, \bibinfo{title}{{KM3-230213A: An Ultra-High Energy Neutrino from a Year-Long Astrophysical Transient},} \doarXiv{2502.12986}

% type= article
\bibitem[{A.~V. Olinto {et~al.}(2021)Olinto {et~al.}}]{POEMMA:2020ykm}
Olinto, A.~V., {et~al.} 2021, \bibinfo{title}{{The POEMMA (Probe of Extreme Multi-Messenger Astrophysics) observatory},} JCAP, 06, 007, \dodoi{10.1088/1475-7516/2021/06/007}

% type= article
\bibitem[{C.~M.~B. Omand {et~al.}(2026)Omand {et~al.}}]{Omand:2026vyl}
Omand, C. M.~B., {et~al.} 2026, \bibinfo{title}{{Multiwavelength Modeling of the Luminous Fast Blue Optical Transient AT2024wpp},} \doarXiv{2601.03372}

% type= article
\bibitem[{A.~N. Otte(2025)Otte}]{Otte:2025dld}
Otte, A.~N. 2025, \bibinfo{title}{{The Trinity-One PeV-Neutrino Telescope},} PoS, ICRC2025, 1136, \dodoi{10.22323/1.501.1136}

% type= article
\bibitem[{D.~A. Perley {et~al.}(2019)Perley {et~al.}}]{Perley:2018oky}
Perley, D.~A., {et~al.} 2019, \bibinfo{title}{{The Fast, Luminous Ultraviolet Transient AT2018cow: Extreme Supernova, or Disruption of a Star by an Intermediate-Mass Black Hole?},} Mon. Not. Roy. Astron. Soc., 484, 1031, \dodoi{10.1093/mnras/sty3420}

% type= article
\bibitem[{D.~A. Perley {et~al.}(2021)Perley {et~al.}}]{Perley:2021ell}
Perley, D.~A., {et~al.} 2021, \bibinfo{title}{{Real-time discovery of AT2020xnd: a fast, luminous ultraviolet transient with minimal radioactive ejecta},} Mon. Not. Roy. Astron. Soc., 508, 5138, \dodoi{10.1093/mnras/stab2785}

% type= article
\bibitem[{D.~A. Perley {et~al.}(2026)Perley {et~al.}}]{Perley:2026clz}
Perley, D.~A., {et~al.} 2026, \bibinfo{title}{{AT2024wpp: An Extremely Luminous Fast Ultraviolet Transient Powered by Accretion onto a Black Hole},} \doarXiv{2601.03337}

% type= article
\bibitem[{S.~J. Prentice {et~al.}(2018)Prentice {et~al.}}]{Prentice:2018qxn}
Prentice, S.~J., {et~al.} 2018, \bibinfo{title}{{The Cow: discovery of a luminous, hot and rapidly evolving transient},} Astrophys. J. Lett., 865, L3, \dodoi{10.3847/2041-8213/aadd90}

% type= article
\bibitem[{L. Sironi \& A. Spitkovsky(2011)Sironi \& Spitkovsky}]{Sironi:2011zf}
Sironi, L., \& Spitkovsky, A. 2011, \bibinfo{title}{{Acceleration of Particles at the Termination Shock of a Relativistic Striped Wind},} Astrophys. J., 741, 39, \dodoi{10.1088/0004-637X/741/1/39}

% type= article
\bibitem[{L. Sironi \& A. Spitkovsky(2014)Sironi \& Spitkovsky}]{Sironi:2014jfa}
Sironi, L., \& Spitkovsky, A. 2014, \bibinfo{title}{{Relativistic Reconnection: an Efficient Source of Non-Thermal Particles},} Astrophys. J. Lett., 783, L21, \dodoi{10.1088/2041-8205/783/1/L21}

% type= article
\bibitem[{V.~A. Villar {et~al.}(2018)Villar, Nicholl, \& Berger}]{Villar:2018toe}
Villar, V.~A., Nicholl, M., \& Berger, E. 2018, \bibinfo{title}{{Superluminous Supernovae in LSST: Rates, Detection Metrics, and Light Curve Modeling},} Astrophys. J., 869, 166, \dodoi{10.3847/1538-4357/aaee6a}

% type= article
\bibitem[{C. Yuan {et~al.}(2025)Yuan, Pfeiffer, Winter, Zaballa, Buson, Testagrossa, \& Azzollini}]{Yuan:2025zwe}
Yuan, C., Pfeiffer, L., Winter, W., {et~al.} 2025, \bibinfo{title}{{An Accretion Flare Interpretation for the Ultra-High-Energy Neutrino Event KM3-230213A},} \doarXiv{2506.21111}

% type= article
\bibitem[{A. Zeolla {et~al.}(2025)Zeolla {et~al.}}]{BEACON:2025qcq}
Zeolla, A., {et~al.} 2025, \bibinfo{title}{{Sensitivity of BEACON to ultra-high energy diffuse and transient neutrinos},} JCAP, 09, 033, \dodoi{10.1088/1475-7516/2025/09/033}

\end{thebibliography}
\bibliographystyle{aasjournalv7}
\appendix
\section{Evolution of energy in pulsar-wind nebula system} 
\label{appsec:energy}
\begin{figure}
\centering
\includegraphics[width=0.48\textwidth]{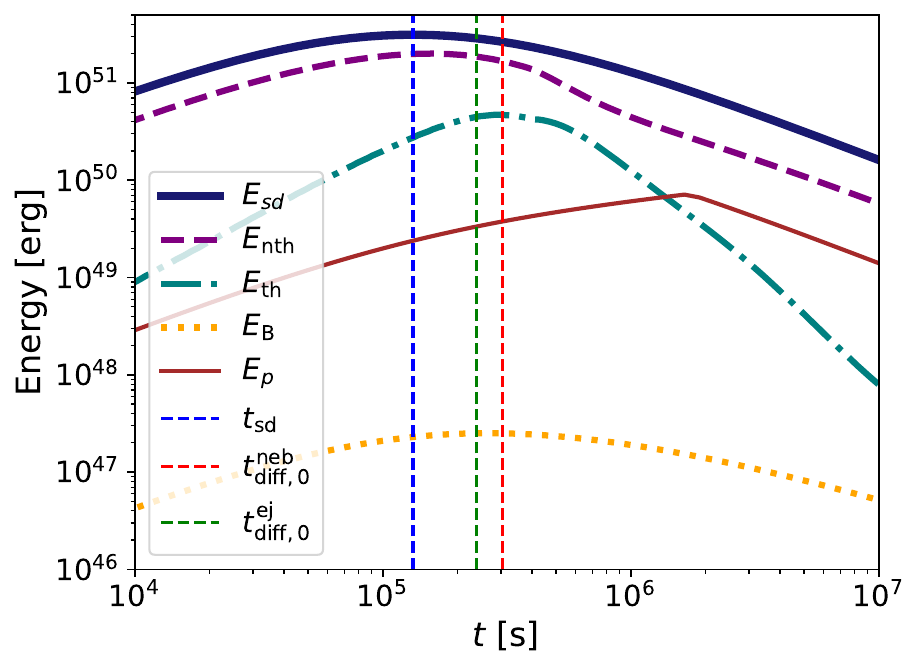}
\caption{\label{fig:energy}Time evolution of the total spindown ($E_{\rm sd}$), non-thermal ($E_{\rm nth}$), thermal ($E_{\rm th}$), magnetic ($E_B$), and CR proton ($E_p$) energy, along with the spindown timescale ($t_{\rm sd}$), and the characteristic nebula ($t_{\rm diff,0}^{\rm neb}$) and ejecta ($t_{\rm diff,0}^{\rm ej}$) diffusion timescales for a chosen parameter set for LFBOTs ($P_i = 1.6$ ms and $B_d = 6 \times 10^{13}$ G).
}
\end{figure}
In this section, we discuss the time evolution of the various energy components of the pulsar-wind-nebula-ejecta system. The spindown energy ($E_{\rm sd}$) of the millisecond pulsar is deposited in the nebula as non-thermal ($E_{\rm nth}$), thermal ($E_{\rm th}$), and magnetic ($E_B$) energy. A part of this energy is also used to do $PdV$ work on the ejecta leading to its expansion. The differential equations governing the evolution of $E_{\rm nth}$, $E_{\rm th}$, and $E_B$ are provided in~\cite{Mukhopadhyay:2024ehs} (see Section 3 and Equations 1,2, and 3 there). However, there is minor difference in the evolution of $E_{\rm th}$. Unlike the case of binary neutron star merger ejecta, the pulsar-powered optical transients resulting from core-collapse, do not have heavy r-process elements in the ejecta. Thus, the heating due to decay of r-process elements which is accounted for by $Q_{\rm rp}^{\rm heat}$ and acts as a source term for $E_{\rm th}$, is not present here. Instead, ordinary SNe and SLSNe, can have nickel in the ejecta, which eventually decay to cobalt and iron. We quantify the rate of heating due to decay of $^{56}$Ni and its contribution to $E_{\rm th}$  using $Q_{\rm heat}^{\rm Ni}$ (see Section~\ref{sec:survey} and ~\citealt{Prepration:2026slsne}).

The time evolution of the various energy components are shown in Figure~\ref{fig:energy} for a LFBOT with $P_i = 1.6$ ms and $B_d = 6 \times 10^{13}$ G. The spindown timescale, $t_{\rm sd} \sim 1.3 \times 10^5$ s where the total energy deposited in the nebula peaks to $\sim 3 \times 10^{51}$ erg. The non-thermal energy roughly follows the evolution of the spindown energy for $t<t_{\rm diff,0}^{\rm neb}$, where $t_{\rm diff,0}^{\rm neb}$ is the characteristic photon diffusion timescale. This is to be expected since $E_{\rm nth} \propto E_{\rm sd}$. However, the non-thermal energy in photons drops rapidly once the nebula becomes transparent $t\gtrsim {\rm a\ few} \times 10^5$ s. Owing to the small ejecta mass, the ejecta diffusion timescale $t_{\rm diff,0}^{\rm ej}$ is $\sim 2 \times 10^5$ s. The thermal photons (and hence $E_{\rm th}$) peak around $t_{\rm diff,0}^{\rm neb}$ because until this time the ejecta reprocesses most of the non-thermal photons in the nebula efficiently. Once the non-thermal photons, a fraction of which act as a source for the thermal photons, can efficiently diffuse out of the nebula (for $t>t_{\rm diff,0}^{\rm neb}$), the thermal energy drops rapidly. The magnetic energy $E_B \propto E_{\rm sd}$ and hence closely follows the evolution of $E_{\rm sd}$. The total energy in CR protons is defined as $E_p = \dot{N}_p \varepsilon_p^{\prime \rm max} t$ (see Equation~\ref{eq:prot_inj_spec}), where $\varepsilon_p^{\prime \rm max}$ is the maximum energy of a CR proton in the comoving frame. Since $\dot{N}_p \propto 1/t$, $\dot{N}_p t$ is a constant. Thus, the evolution of $E_p$ depends on $\varepsilon_p^{\prime \rm max}$. We discuss the evolution of $\varepsilon_p^{\prime \rm max}$ in Appendix~\ref{appsec:proton_details} and show it in Figure~\ref{fig:epcut}, which readily explains the evolution of $E_p$ in Figure~\ref{fig:energy}.
\section{CR proton injection and acceleration}
\label{appsec:proton_details}
\begin{figure}
\centering
\includegraphics[width=0.48\textwidth]{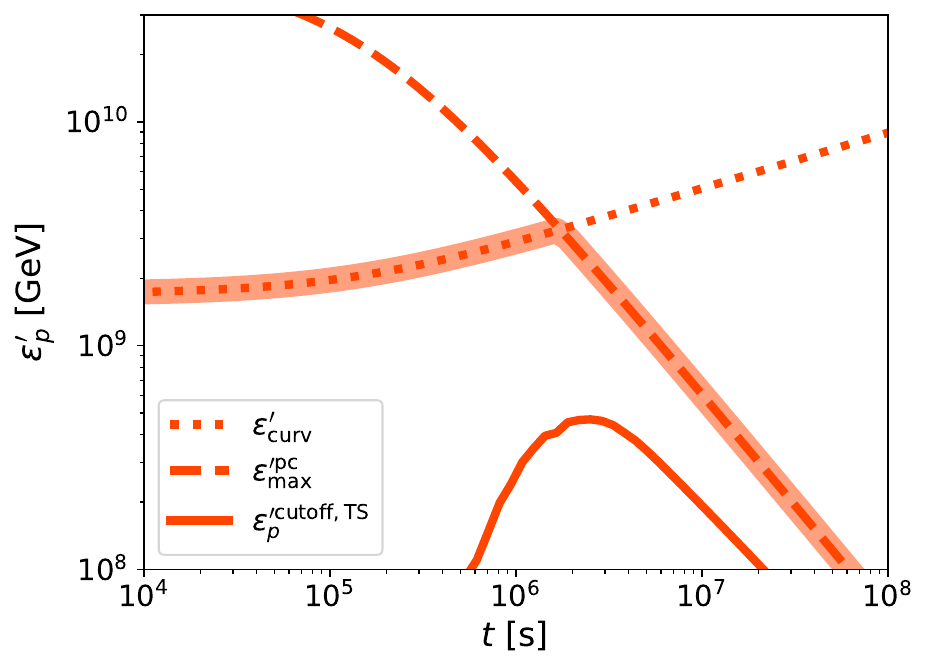}
\caption{\label{fig:epcut}Time evolution of the CR proton energy in the polar cap and TS regions, where $\varepsilon_{\rm max}^{\prime \rm pc}$ is defined as the maximum energy that the protons can be accelerated to in the polar cap region, $\varepsilon_p^{\prime \rm cutoff, TS}$ is the cut off energy in the TS region, the proton energy limited by curvature losses is given by $\varepsilon^\prime_{\rm curv}$. The final cutoff energy for the CR protons in a LFBOT ($\varepsilon_p^{\prime \rm max}$) with $P_i = 1.6$ ms and $B_d = 6 \times 10^{13}$ G, is denoted by the shaded band.
}
\end{figure}
In this section we discuss the injection spectrum of the CR protons, followed by their acceleration in the polar cap and TS region, and their resulting maximum energy. 

The injection spectrum for the protons is normalized using the Goldreich-Julian charge density ($n_{\rm GJ}$). The spectrum is modeled using a one-shot acceleration across the potential difference created in the vicinity of the polar cap
\begin{align}
\label{eq:prot_inj_spec}
\frac{d\dot{N}_{p,\rm inj}}{d \varepsilon_p^\prime } &= \dot{N}_p\delta\big(\varepsilon_p^\prime -  \varepsilon_p^{\prime \rm cutoff, pc} \big)\,,\nonumber\\
\dot{N}_p &= n_{\rm GJ}2 A_{\rm pc} c = \frac{4 \pi^2}{e} \frac{R_*^3}{c} \frac{B_d}{P^2}\,,
\end{align}
where the primed quantities correspond to the comoving frame, $A_{\rm pc}$ is the size of the polar cap given by $A_{\rm pc} = \pi R_*^2 (R_*/R_{\rm lc})$, the light cylinder radius $R_{\rm lc} = c P/(2 \pi)$, the spin period $P=P_i \big( 1+t/t_{\rm sd} \big)^{1/2}$. The cut off energy in the comoving frame associated with the polar cap acceleration is determined by $\varepsilon_p^{\prime \rm cutoff, pc} = {\rm  min} \left[\varepsilon^{\prime  \rm pc}_{\rm max}, \varepsilon_{\rm curv}^\prime  \right]$. The protons are accelerated by the potential difference created at the polar cap for which the maximum energy that can be achieved is given by $\varepsilon^{\prime \rm pc}_{\rm max}= 4 \eta_{\rm gap} e B_d \left( \pi R_*/(c P) \right)^2 R_*$. We assume that the charged particle experiences a fraction $\eta_{\rm gap} = 0.3$ of the potential gap. The curvature energy loss for the same particles can be calculated using $\varepsilon_{\rm curv}^\prime = \left( 3 m_p^4 c^8 B_d R_{\rm curv}^2/(2 e)\right)^{1/4}$. The radius of curvature $R_{\rm curv} = R_{\rm lc}$. The results for the maximum energy across the polar cap and TS region for a LFBOT with $P_i = 1.6$ ms and $B_d = 6 \times 10^{13}$ G is shown in Figure~\ref{fig:epcut}. We note that for the chosen parameter set, the TS acceleration does not play a role and the maximum energy for the CR protons ($\varepsilon^{\prime}_{p, \rm max}$) is given by the polar cap acceleration. For $t \lesssim 10^6$ s, the curvature losses govern $\varepsilon^{\prime}_{p, \rm max}$. For $t>10^6$ s, $\varepsilon^{\prime}_{p, \rm max} = \varepsilon^{\prime  \rm pc}_{\rm max}$ which falls off as $1/t$. This explains the rapid fall off in the neutrino energy fluence for $t \gtrsim 2 \times 10^6$ s, as seen in Figure~\ref{fig:nu}.

\begin{figure}
\centering
\includegraphics[width=0.32\textwidth]{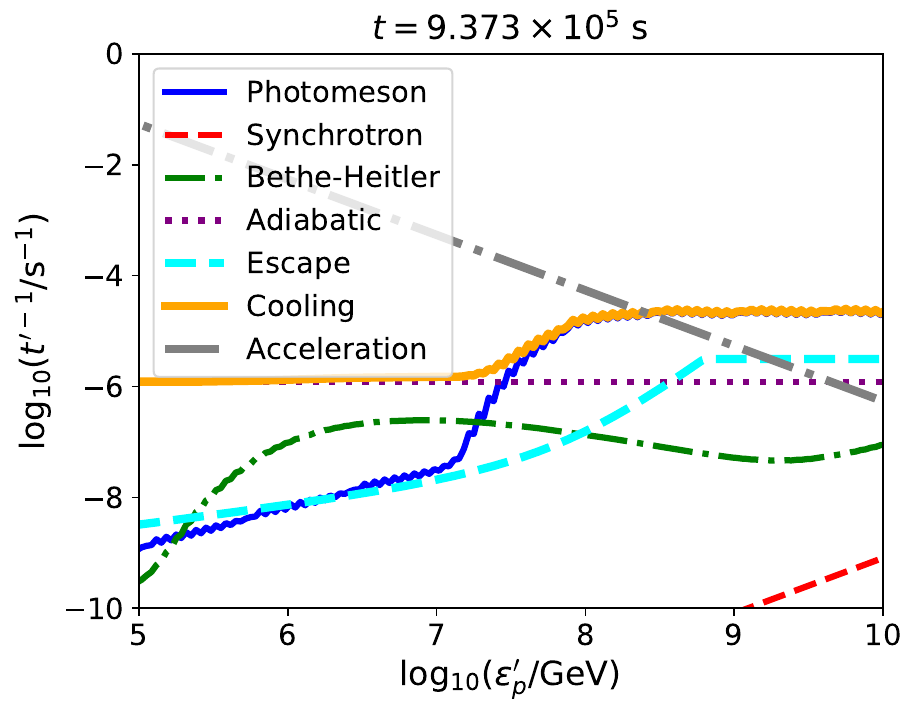}
\includegraphics[width=0.32\textwidth]{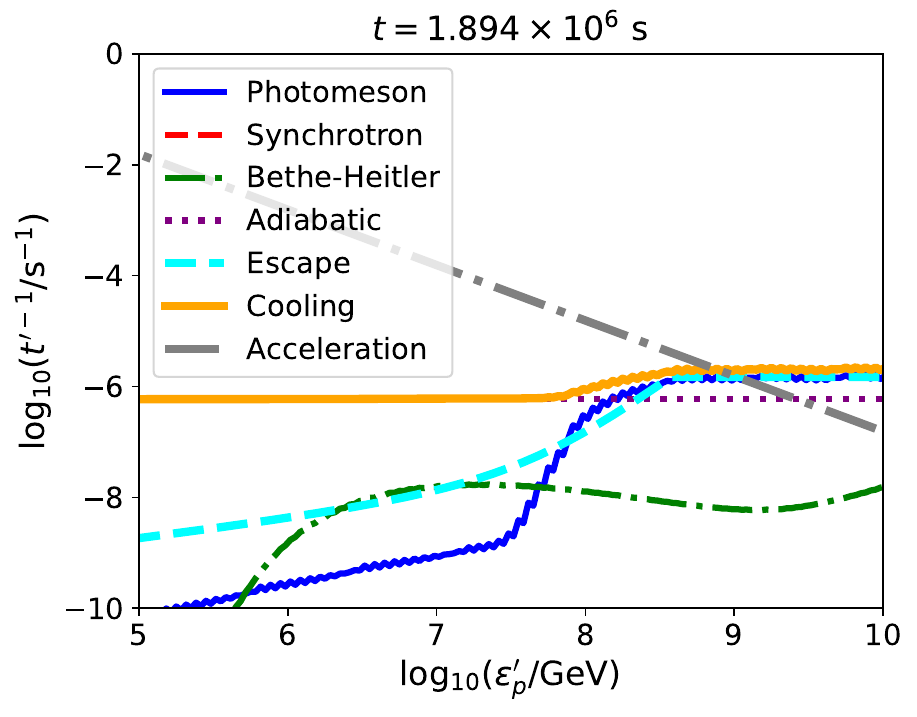}
\includegraphics[width=0.32\textwidth]{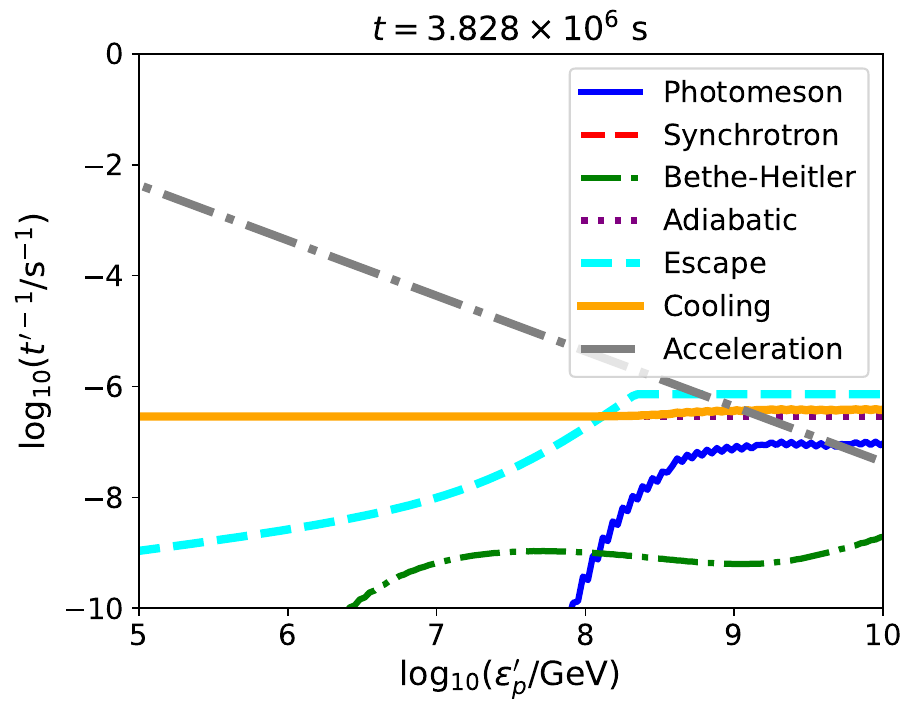}
\caption{\label{fig:nu_timescales}The rates of various acceleration and cooling processes with comoving proton energy $\varepsilon_p^\prime$ at various timesnaps for a chosen parameter set for LFBOTs ($P_i = 1.6$ ms and $B_d = 6 \times 10^{13}$ G).
}
\end{figure}
A fraction of the protons can be re-accelerated at the termination shock (TS) region. The maximum energy that the protons can be accelerated to in this region is computed by balancing the acceleration ($t_{\rm acc}^\prime$)  and loss ($t_{\rm loss}^\prime$) timescales. The acceleration timescale is given by $t_{\rm acc}^{\prime} = \eta_{\rm acc}\varepsilon_p^\prime/\big(e c B_{\rm neb}^\prime \big)$, where $B_{\rm neb}^\prime = \sqrt{2}\left(E_B/(R^2 \Gamma_{\rm ej}^2 \beta_{\rm ej} c t) \right)^{1/2}$ is the co-moving magnetic field and the efficiency of acceleration is chosen to be $\eta_{\rm acc} = 1$. The loss rate is given by $t_{\rm loss}^{\prime -1}=t_{\rm esc}^{\prime  -1}+t_{\rm cool}^{\prime -1}$, where $t_{\rm esc}^\prime $ and $t_{\rm cool}^\prime $ are the escape and the cooling timescales for the protons, respectively. The escape timescale for the protons is given by $t_{\rm esc}^\prime  = {\rm max}\left[ R(t)^2/D_c(\varepsilon_p^\prime), R(t)/c \right]$, where for the first term we assume a Kolmogorov cascade for the diffusion of the CR protons~\citep{Harari:2013pea,Mukhopadhyay:2024ehs} and the second term is the light crossing time. The cooling rate for the CR protons is given by $t_{\rm cool}^{\prime -1} =  t_{\rm p\gamma}^{\prime -1} + t_{\rm sync}^{\prime -1} + t_{\rm BH}^{\prime -1} + t_{\rm dyn}^{\prime -1}$, where $t_{pp}^\prime , t_{\rm p\gamma}^\prime , t_{\rm sync}^\prime , t_{\rm BH}^\prime , t_{\rm dyn}^\prime $ denote $pp$, $p\gamma$, proton synchrotron (sync), Bethe-Heitler (BH), and dynamical (dyn) cooling timescales respectively. The relevant rates for various processes are shown in Figure~\ref{fig:nu_timescales}. At $t\sim 9 \times 10^5$ s, CR protons with $\varepsilon_p^\prime>10^8$ GeV are efficiently cooled through the photomeson channel relevant for neutrino production. With increase in time, the photomeson production efficiency decreases for similar values of $\varepsilon_p^\prime$. For $t\gtrsim 2 \times 10^6$ s the CR protons mostly escape.
\section{Estimating the diffuse flux}
\label{appsec:diffuse_comp}
This section deals with estimating the resultant diffuse neutrino flux from a population of LFBOTs. The diffuse neutrino contribution from LFBOTs as sources can be computed using
\be
\label{eq:diffuse}
\phi_\nu = \int_0^{z_{\rm max}} dz \frac{dV_c}{dz d\Omega} \int_{M_{\rm AB}^{\rm min}}^{M_{\rm AB}^{\rm max}} dM_{\rm AB} \Phi (z, M_{\rm AB}) \frac{(1+z)}{4\pi d_L(z)^2} \frac{dN}{dE_\nu}\big((1+z)E_\nu, B_d, P_i | M_{\rm AB} \big)\,,
\ee
where the differential comoving volume element per unit redshift and unit solid angle $dV_c/(dz\ d\Omega) =  \big(c/H(z)\big) d_c(z)^2$, the comoving distance ($d_c(z)$) is related to luminosity distance $d_L = (1+z)d_c$, the Hubble expansion rate for flat $\Lambda$CDM cosmology at a given redshift $z$ is given by $H(z) = H_0\sqrt{\Omega_m (1+z)^3 + \Omega_\Lambda}$, the matter and vacuum energy density fraction is given by $\Omega_m = 0.3$ and $\Omega_\Lambda = 0.7$ respectively. The luminosity function for LFBOTs is given by $\Phi (z, M_{\rm AB})$. Since the luminosity function for LFBOTs is not well-defined we adopt a simple phenomenological form spanning the observed magnitude ranges for LFBOTs~\citep{Perley:2018oky,Perley:2021ell} along with a declining bright-end tail. Thus, we have
\be
\Phi (z, M_{\rm AB}) = \dot{\rho}_0 \frac{\psi_z}{\psi_0}\begin{cases}
0,\hspace{0.5cm} M_{\rm AB} > - 20\,,\\
1,\hspace{0.5cm} -22 \leq M_{\rm AB} \leq -20\,,\\
\exp \left( (M_{\rm AB} - 22)/0.5  \right),\hspace{0.1cm} M_{\rm AB}<-22\,,
\end{cases}\,
\ee
where the local ($z=0$) rate density is given by $\dot{\rho}_0 = \mathcal{N}\dot{\rho}_0^{\rm LFBOTs}$, where $\mathcal{N}$ is a normalization constant. The redshift dependent star formation rate (SFR) is given by~\citep{Madau:2014bja} (see equation 15 there)
\be
\psi_z (z) = 0.015 \times \frac{(1+z)^{2.7}}{1 + \left( (1+z)/2.9 \right)^{5.6} }\,,
\ee
and $\psi_0 = \psi_z (z = 0)$ gives the local SFR. The energy in the observer frame ($E_\nu$) is related to the source frame energy ($E_\nu^\prime$) by $E_\nu^\prime = (1+z)E_\nu$. The number per energy bin $dN/dE_\nu$ is computed as a function of $E_\nu^\prime$, where we relate the optical luminosity ($M_{\rm AB}$) of LFBOTs to the pulsar parameters ($P_i,B_d$). The integrands are integrated over redshifts $0$ to $z_{\rm max} = 4$ and the absolute AB magnitude $M_{\rm AB}$ where we choose $M_{\rm AB}^{\rm min} = -24$ and $M_{\rm AB}^{\rm max} = -20$.

The term  $dN/dE_\nu \big((1+z)E_\nu, B_d, P_i | M_{\rm AB} \big)$ relates the peak $M_{\rm AB}$ luminosity for LFBOTs in the $r-$band for thermal photons to the two crucial central engine parameters initial spin period ($P_i$) and dipolar magnetic field strength ($B_d$). The algorithm we use to do this is as follows. We define a grid of $ M_{\rm AB}$ ranging from $M_{\rm AB}^{\rm min}$ to $M_{\rm AB}^{\rm max}$ consisting of $20$ points. For each value of $M_{\rm AB}^j \forall\ j \in \{1,20\}$ in the grid, we plot the corresponding contour in the $B_d - P_i$ plane. We overlay the contour of $t_- = 6$ days corresponding to the fall time of LFBOTs. We then pick the value of $B_d$ and $P_i$ where the two contours intersect and choose that as the parameter set for the central engine for the chosen $M_{\rm AB}^j$. The motivation behind choosing the intersection of the $M_{\rm AB}$ and $t_-$ contours is the fact that LFBOTs have a rapidly evolving lightcurve. Hence, the intersection between the contours allows us to select the parameters corresponding to a population of LFBOTs that satisfy the observed features appropriately. In case there are multiple intersection points (in scenarios where this happens, the intersection points are very close to each other), we choose the lowest value of $P_i$ and $B_d$ where the two contours intersect. We then evaluate $dN/dE_\nu$ for each set of $(B_d,P_i)$ chosen corresponding to $M_{\rm AB}^j$ to use for the integration in Eq.~\eqref{eq:diffuse} to compute the diffuse flux.
\end{document}